\documentclass[12pt]{article}
\usepackage{xr}
\usepackage{amsmath}
\usepackage{amssymb}
\usepackage[symbol]{footmisc}
\usepackage{amsfonts}
\usepackage{dsfont}
\usepackage{listings}
\usepackage{xcolor}
\usepackage{url}
\usepackage{mathtools}
\usepackage{hyperref}
\usepackage[colorinlistoftodos]{todonotes}
\usepackage{colortbl}
\usepackage{graphicx}
\usepackage{ulem}
\usepackage{setspace} 
\usepackage{color,soul}
\usepackage{comment}
\usepackage{multirow}
\usepackage{xargs} % Use more than one optional parameter in a new commands

\usepackage{listings}
\lstdefinestyle{Rstyle}{
  language=R,
  basicstyle=\ttfamily\small,
  breaklines=true,
  keywordstyle=\color{blue},
  commentstyle=\color{green!60!black},
  identifierstyle=\color{black},
  backgroundcolor=\color{gray!10!white},
  stringstyle=\color{red}
}

\newcommandx{\jcq}[2][1=]{\todo[inline,linecolor=blue,backgroundcolor=blue!25,bordercolor=blue,#1]{#2}}
\renewcommandx{\jcq}[2][1=]{} % pour retirer les commentaires
\def\checkmark{\tikz\fill[scale=0.4](0,.35) -- (.25,0) -- (1,.7) -- (.25,.15) -- cycle;}

\usepackage{cancel}

\newcommand{\btheta}{\boldsymbol{\theta}}

\begin{document}
% \listoftodos

%\Cor{}{}{Reste des JdS})

\begin{center}
{\Large
	{\sc %Number of zero velocity points in oscillatory models of handwriting as a relevant criteria for dysgraphia diagnostic\\

    A statistical procedure to assist dysgraphia detection through dynamic modelling of handwriting
	}
}
\bigskip

\underline{Yunjiao Lu}$^{1,2}$\footnote{Corresponding author information: Yunjiao Lu, Institute of Cancer Research, 237 Fulham Broadway, SW3 6JB, London, United-Kingdom (e-mail: \url{yunjiao.lu@icr.ac.uk}).} \& Jean-Charles Quinton$^{1}$ \& Caroline Jolly$^{3}$ \& Vincent Brault$^{1}$
\bigskip

{\it
$^{1}$ Univ. Grenoble Alpes, CNRS, Grenoble INP\footnote{Institute of Engineering Univ. Grenoble Alpes}, LJK, 38000 Grenoble, France\\

$^{2}$ Institute of Cancer Research, SW3 6JB London, United-Kingdom\\

$^{3}$ CNRS, University Grenoble Alpes, University Savoie Mont Blanc, LPNC, 38000 Grenoble, France

}
\end{center}
\bigskip

\section*{Abstract}
Dysgraphia is a neurodevelopmental condition in which children encounter difficulties in handwriting. Dysgraphia is not a disorder per se, but is secondary to neurodevelopmental disorders, mainly dyslexia, Developmental Coordination Disorder (DCD, also known as dyspraxia) or Attention Deficit Hyperactivity Disorder (ADHD). Since the mastering of handwriting is central for the further acquisition of other skills such as orthograph or syntax, an early diagnosis and handling of dysgraphia is thus essential for the academic success of children. In this paper, we investigated a large handwriting database composed of 36 individual symbols (26 isolated letters of the Latin alphabet written in cursive and the 10 digits) written by 545 children from 6,5 to 16 years old, among which 66 displayed dysgraphia (around 12\%). To better understand the dynamics of handwriting, mathematical models of nonpathological handwriting have been proposed, assuming oscillatory and fluid generation of strokes (Parsimonious Oscillatory Model of Handwriting [ (André, 2014)]). The purpose of this work is to study how such models behave when applied to children dysgraphic handwriting, and whether a lack of fit may help in the diagnosis, using a two-layer classification procedure with different compositions of classification algorithms.

\paragraph{Keywords:}Statistical modelling, supervised classification, dysgraphia detection, Parsimonious Oscillatory Model of Handwriting

\paragraph{Data availability statement:} The raw data that support the findings of this study are not available. Processed data and scripts are openly available in GitLab: \url{https://gricad-gitlab.univ-grenoble-alpes.fr/braultv/ecriture_mse3}

\section{Introduction}

%\jcq{Désolé pour le compactage/reformatge des références entre parenthèses; si non accepté/standard pour les JdS, n'hésitez pas à reverter. Ce ne sont pas non plus les normes APA qui sont appliquées, sans quoi le format risque aussi de surprendre Caroline.}
%Learning to handwriting is a long and important process for successful integration into society until adulthood, but some children, called \textit{dysgraphics}, may have acquisition problems. To detect children with dysgraphia, the BHK test (\textit{Brave Handwriting Kinder}), currently in use, consists of copying a text by hand for 5 minutes and is evaluated by professionals (see for example Hamstra-Bletz et al. (1987) and the french adaptation Charles et al. (2004)) but it is both long and subjective. To propose a diagnostic aid, some authors study the dynamic parameters collected by having the children write on tablets (see for example Asselborn et al. (2018), Deschamps et al. (2019) or Devillaine et al. (2020)): the principle is to take dynamic parameters (such as the time to make certain strokes) and see how they can help predict dysgraphia.
Handwriting is crucial for a successful integration into society, yet hinges on a learning process that usually spans over many years \cite{fancher2018handwriting}. Handwriting deficits ---called dysgraphias--- may therefore yield severe consequences spanning from infancy to adulthood \cite{mccloskey2019developmental, chung2020disorder}.
%In France, dysgraphia is usually diagnosed using the \textit{Concise Evaluation Scale for Children's Handwriting} (aka BHK test),
In France, dysgraphia is usually diagnosed using the BHK test,
which consists in copying a text for 5 minutes, with the handwritten production then rated on 13 qualitative criteria and one speed criterion by psychomotricity specialists (adapted in French by Charles et al. (2004) based on \cite{hamstra1987concise}). Quality criteria include inconsistencies in size, vertical or horizontal misalignments, letter collisions and distortions, interrupted joins between letters, corrections, hesitations and unsteady trace, or chaotic handwriting. Evaluating such criteria solely based on visual traces is difficult, as many of these criteria rely on dynamical aspects of handwriting production. The use of tablets to record and exploit the handwriting dynamics via different kind of machine learning or deep learning methods has been proposed (\cite{jolly2023dysgraphia, asselborn2018automated,deschamps2019methodological,devillaine2021analysis} and reviewed \cite{danna2023tools}.

In this article, we investigated a large handwriting database composed of 36 individual symbols (the 26 isolated letters of the latin alphabet written in cursive and the 10 digits) written by 545 children from 6,5 to 16 years old, among which 66 displayed dysgraphia (around 12\%). The study from which the database was collected was conducted in accordance with the Helsinki Declaration and was approved by the Grenoble University Ethics Committee (CERGA, agreement 2016-01-05-79). 
% \todo[inline, color=blue!40]{YJ: Present our data and other papers which already studied these data and how our method different from theirs}

%On the other hand, models of handwriting have been proposed by different authors (see Hollerbach (1984) and André et al. (2014)). The purpose of this work is to study how these models, based on the assumption of fluent handwriting, behave when applied to model the handwriting of children diagnosed as dysgraphic and to study how the results can help in the diagnosis of the latter. To this end, the Parsimonious Oscillatory Model of Handwriting (POMH) and the method of parameter estimation are first presented and then a study on the first differences between the modeling on children diagnosed as dysgraphic or not is proposed. A discussion on the use of these results concludes this work.
To better understand and exploit its dynamics, mathematical models of non-pathological handwriting have been proposed, assuming oscillatory and fluid generation of strokes (\cite{andre2014parsimonious,hollerbach1981oscillation}). The purpose of this work is to study how such models behave when applied to dysgraphic handwriting, and whether a lack of fit may help in the diagnosis. To this end, the Parsimonious Oscillatory Model of Handwriting (POMH; \cite{andre2014parsimonious}) and the associated method for parameter estimation are first presented. 

Then, a first study on the differences in model estimation between dysgraphic and typically developing (TD) children is presented. In sections \ref{sec:pomh} and \ref{sec:transi_pomh_classif}, we explain the practical considerations and important parameters rising up when we apply POMH on the handwriting data. Based on the conception of POMH, the assumption for the model to successfully reconstruct a handwriting trace requires the handwriting motion to be fluent. In practice, due to the noise in the handwriting signal, a closing operator with structural element of size $w$ should be applied on the raw velocity data, to obtain the moments of zero velocity for POMH reconstruction. It is essential to choose a "good" $w$ value to make the reconstruction error (or distance) of POMH discriminative between individuals having dysgraphia and those without notable writing problems. It is also shown by our studies that age is another important factor to consider in determining the number of zero velocities. In sections~\ref{sec:classif}~and~\ref{sec:res} we exploit the features extracted from POMH model by a two-layer classification procedure with different compositions of classification algorithms, i.e. General Linear Model, Random Forest, Support Vector Machine and Counting. Through Cross Validation, the results of classifiers with different compositions are evaluated. A discussion on the future use of our results concludes this work.

\section{Parsimonious Oscillatory Model of Handwriting}\label{sec:pomh}

The Parsimonious Oscillatory Model of Handwriting (POMH) introduced in \cite{andre2014parsimonious} assumes that handwriting can be approximated by two orthogonal oscillators, corresponding to different degrees of freedom in hand movements (e.g., movement of the wrist and extension of fingers grasping the pen). While this also applies to earlier models of handwritten trajectories, POMH additionally makes equations symmetrical (e.g., compared to \cite{hollerbach1981oscillation}), allowing to model handwriting movements independently of movement orientation and direction. Particularly, oscillators controlled by the human writer may not align with the axes ($x$ and $y$) of the tablet used for recording handwriting trajectories (e.g., due to slanted writing, or left-handed writers twisting their wrists or the tablet). Nevertheless, the approximation and assumptions underpinning our developments do not depend on such alignment, so that the natural axes of the tablet will be used in the rest of the paper. The model is defined as follows:
\[\left\{   \begin{array}{l}
        \dot{x}(t) = a_x(t)\sin(\omega_x(t)\times t + \phi_x(t))\\
        \dot{y}(t) = a_y(t)\sin(\omega_y(t)\times t + \phi_y(t))
    \end{array}\right.\]
%where $\dot{x}$ (resp. $\dot{y}$) is the velocity on the axis $x$ (resp. $y$) and $a_x$, $\omega_x$, $\phi_x$, $a_y$, $\omega_y$, and $\phi_y$ are constant-by-part functions such that there exists $t_{x,1}, \cdots, t_{x,N_x}$ the instants of zero velocity in $x$ axis (resp. $y$) such that for all $i\in\{1,\ldots,N_x-1\}$ and $t\in[t_{x,i};t_{x,i+1}[$:
where $\dot{x}$ (resp. $\dot{y}$) is the velocity on the $x$-axis (resp. $y$) and $a_x$, $\omega_x$ and $\phi_x$ (resp. $a_y$, $\omega_y$, $\phi_y$) are piecewise constant functions. The associated break points of the model $t_{x,1}, \cdots, t_{x,N_x}$ correspond to the instants of zero velocity along the $x$-axis (with the equivalent parametrization for the $y$-axis), such that for all $i\in\{1,\ldots,N_x-1\}$ and $t\in[t_{x,i};t_{x,i+1}[$:
\begin{eqnarray*}
     %&&\omega_x(t)=\frac{\pi}{t_{x,i+1}-t_{x,i}},\quad \phi_x(t)=-\frac{\pi t_{x,i}}{t_{x,i+1}-t_{x,i}},\quad a_x(t)=\frac{\pi}{2\left(t_{x,i+1}-t_{x,i}\right)}\sum_{j|t_j\in[t_{x,i};t_{x,i+1}[}\frac{x_{t_{j+1}}-x_{t_{j}}}{t_{j+1}-t_{j}}\\
     &&\omega_x(t)=\frac{\pi}{t_{x,i+1}-t_{x,i}},\quad \phi_x(t)=-\frac{\pi t_{x,i}}{t_{x,i+1}-t_{x,i}}\\
     \text{and}&&a_x(t)=\frac{\pi}{2\left(t_{x,i+1}-t_{x,i}\right)}\sum_{j|t_j\in[t_{x,i};t_{x,i+1}[}\frac{x_{j+1}-x_{j}}{t_{j+1}-t_{j}}\\
     %\text{and}&&a_x(t)=\frac{\pi}{2\left(t_{x,i+1}-t_{x,i}\right)}\sum_{j|t_{x,j}\in[t_{x,i};t_{x,i+1}[}\frac{x_{t_{x,j+1}}-x_{t_{x,j}}}{t_{x,j+1}-t_{x,j}}\\
\end{eqnarray*}
%with the symmetric parametrization for the $y$ axis.The estimation of the instants of zero velocity, i.e. the break points of the model, is therefore crucial for the reconstruction of the letters.
with $(t_j,x_j,y_j)$ the recorded timestamps and coordinates of the pen point from the tablet. The correct estimation of the instants of zero velocity is therefore crucial for the reconstruction of the letters.
\subsection{Signal preprocessing with mathematical morphological operator "closing" with size $w$}
%\VB{Plutôt les instants 0 puisqu'on a dit avant que les coefficients dépendaient de ceux ci. On peut peut-être mettre une image non ?}
To estimate the coefficients ($\alpha_x, \omega_x, \phi_x, \alpha_y, \omega_y, \phi_y$) in POMH which are constant-by-part, the instants of zero velocity, i.e. the break points of the model should preliminarily be estimated.
\begin{figure}[!ht]
    \centering
    \includegraphics[width = \textwidth]{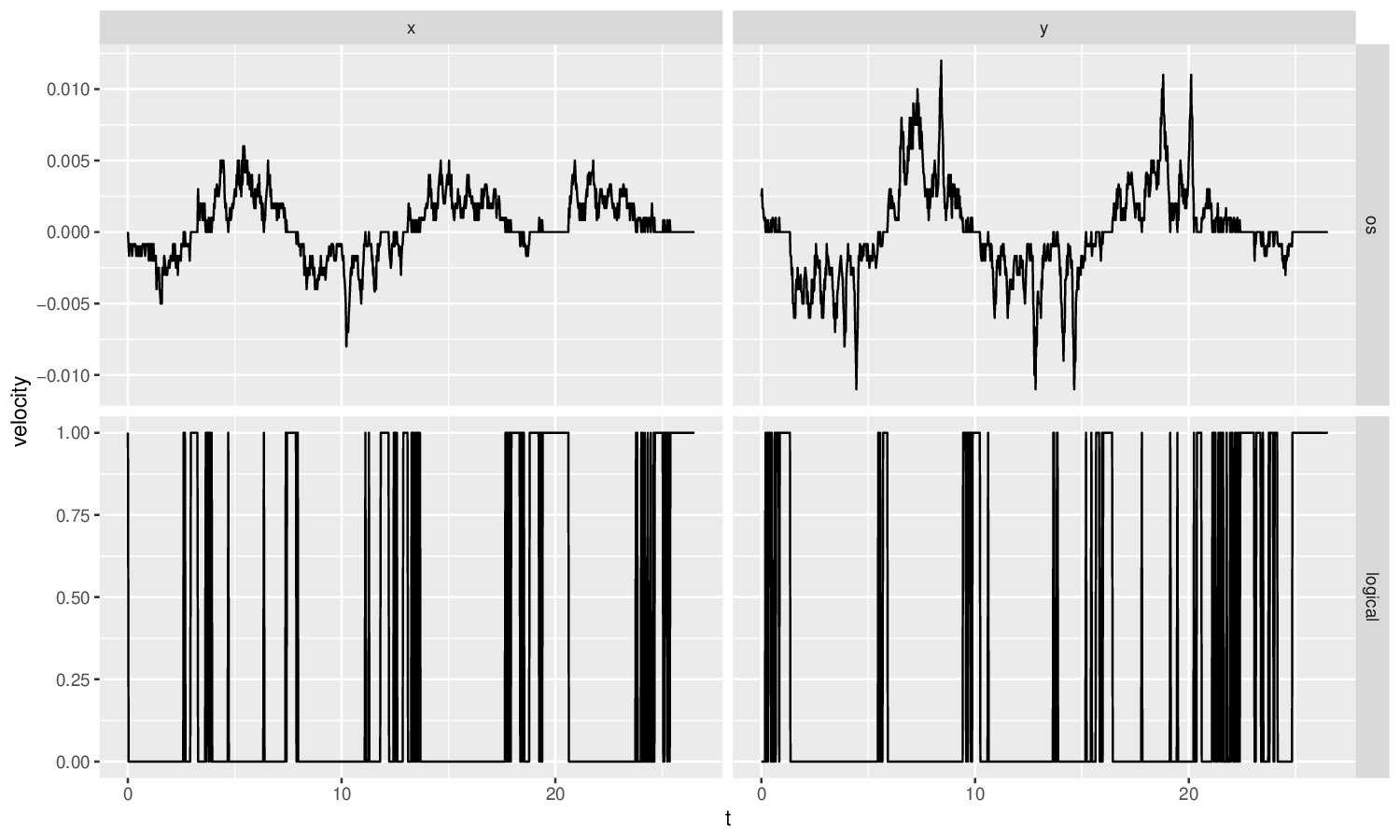}
    \caption{The original signal (os) of the velocity and the logical velocity on both axis.}
    \label{fig:v_og+logi_os}
\end{figure}
\begin{figure}[!ht]
    \centering
    \includegraphics[width = \textwidth]{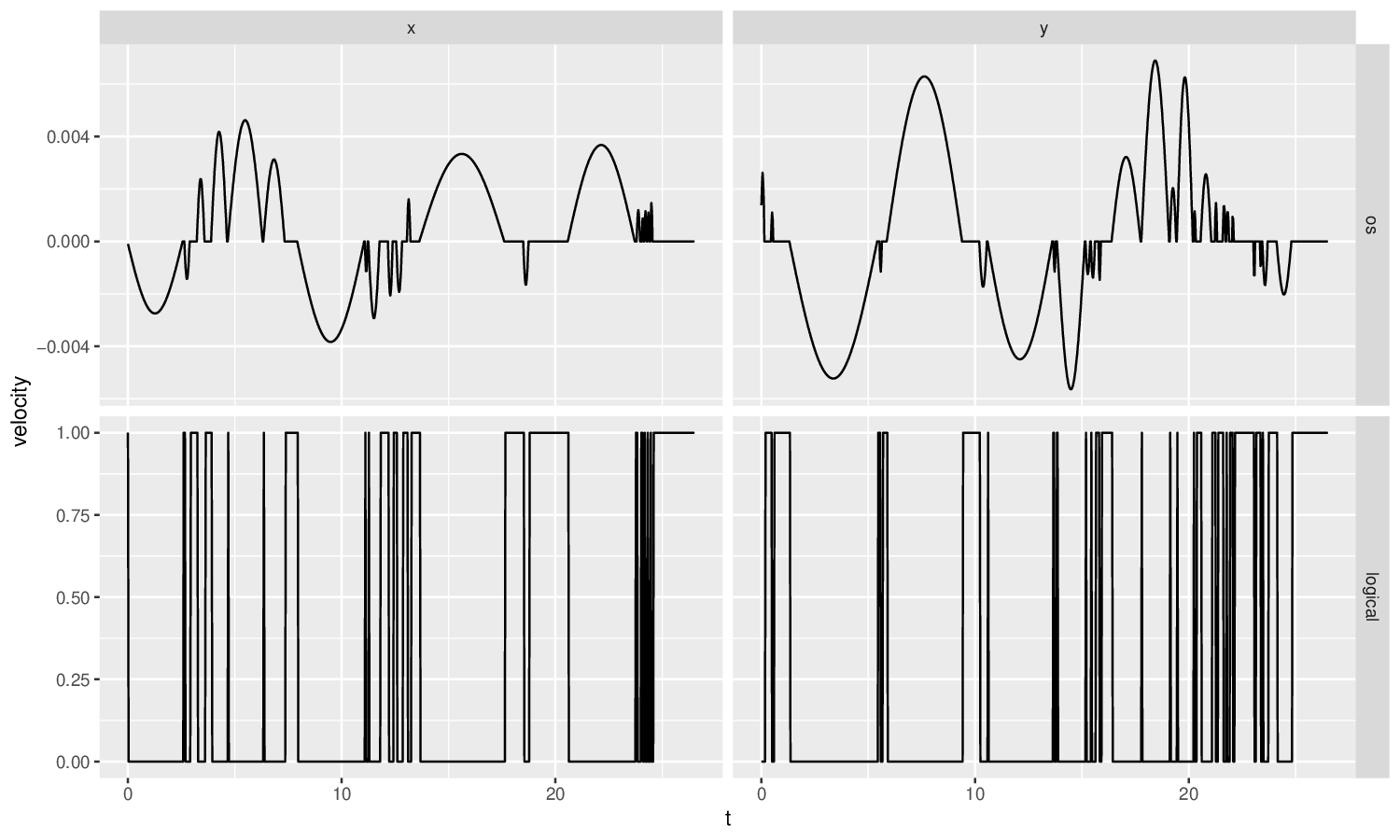}
    \caption{Signal applying closing operator $w=3$ and the corresponding logical velocity for both axis.}
    \label{fig:v_os+log_w3}
\end{figure}
\begin{figure}[!ht]
    \centering
    \includegraphics[width = \textwidth]{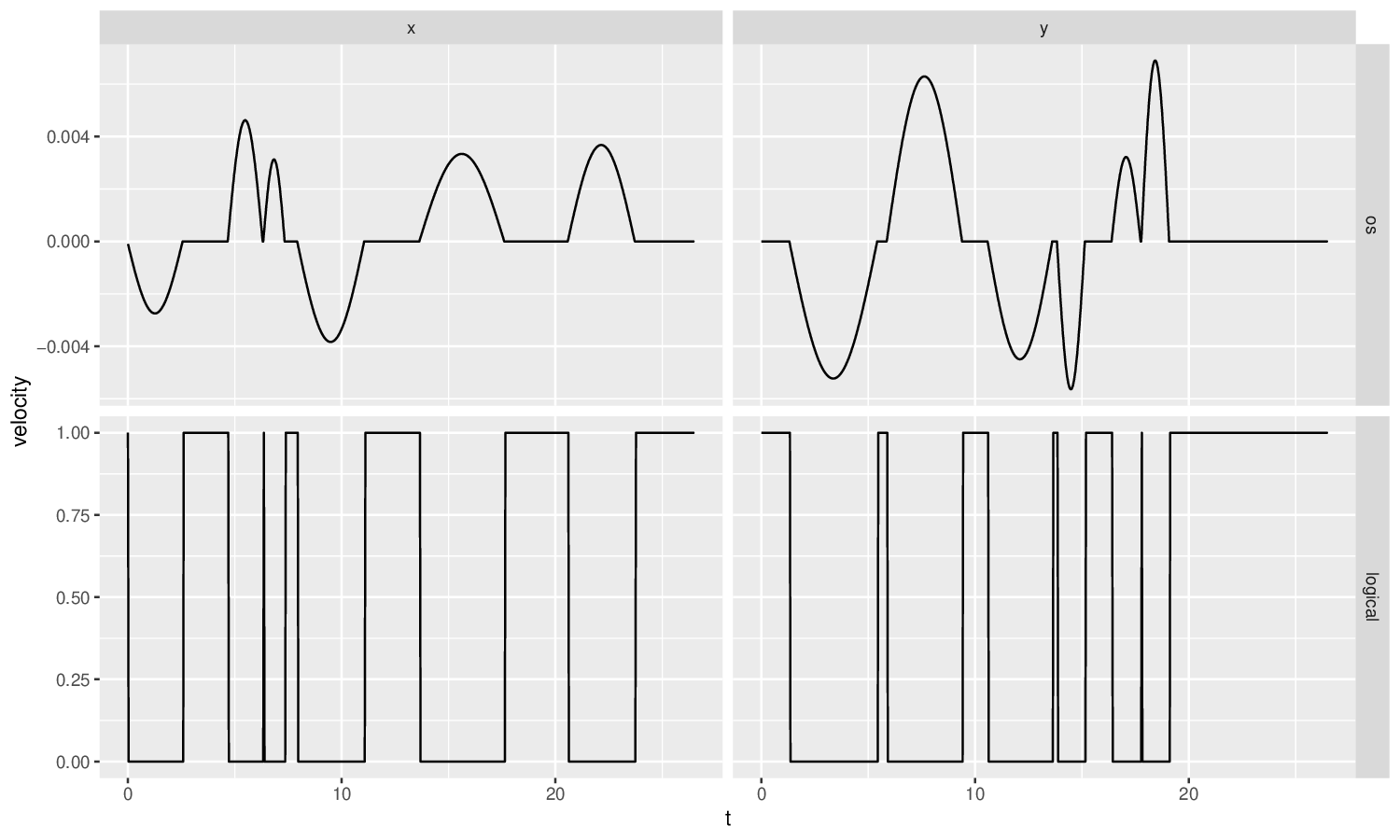}
    \caption{Signal applying closing operator $w=35$ and the corresponding logical velocity for both axis.}
    \label{fig:v_os+log_w35}
\end{figure}

Our data are composed of isolated symbols (letters or digits). We have the position of the pen $(x_t)$ and $(y_t)$ at a regular frequency of $\simeq200$Hz. The moments of zero velocity can be obtained from the recorded handwriting data by calculating the average speed between $t_i$ and $t_{i+1}$. However, the signals are noisy (Fig.~\ref{fig:v_og+logi_os} original signal(os)) and must be filtered to find a meaningful number of zero velocities in the dynamics of the handwriting of a symbol. To do so, the closing morphological operator (acting as a low-pass filter) is applied on a logical variable indicating whether the speed is cancelled (Fig.~\ref{fig:v_og+logi_os} logical). 
%\jcq{Ci-dessus j'ai simplement tenté de plus m'appuyer sur les morceaux ajoutés en section précédente, et clarifier sur quoi le filtre est appliqué (i.e. séparément en $x$ et $y$, et sur des symboles isolés.}
%\yjl{Do you mean signlas of individual symbols by "discretized signals ? If so I suggest "signlas of individual" instead of "discretizd signals" }

 Mathematical Morphology (MM) is a theory for the analysis of spatial structure. It is first designed for binary image processing. The two fundamental MM operators are dilation and erosion which can be seen as a convolution of the original binary image $A$ and the structural element $B$. The structural element, itself a binary image, usually a disk or a square, performs as a probe inside the image $A$ and concludes whether the structure in $A$ fits or misses the shape of $B$. The other two basic operators in MM are opening and closing. Opening is erosion followed by dilation. Closing is dilation followed by erosion. In a binary image, the opening operator removes isolated pixels with value 1 and the closing operator closes the holes (isolated pixels with value 0). We set the binary vector as 1 when $v_x = 0$ and 0 when $v_x \neq 0$. The structural element is a vector of 1 with length $w$ where $w$ is an odd number. In our case the closing operator is suitable to identify the zero velocity time interval by removing the isolated non zero instants in the interval (Fig.~\ref{fig:v_os+log_w3}~and~\ref{fig:v_os+log_w35}). The $w$ values range from 3 to $w_\text{max}$ by an interval of 2. The number of zero velocities decreases when the value $w$ increases. The parameter $w_\text{max}$ is the maximum value that $w$ can take, in other words, $3 \leq w \leq w_\text{max}$ and the value is chosen empirically. In the end of Section~\ref{sec:transi_pomh_classif}, it is explained how to estimate $w$ and it is obvious that the estimator $\hat{w}$ is dependent on the value of $w_\text{max}$.  We aim at choosing the optimal $w_\text{max}$, in the sens that it gives the estimator $\hat{w}$ which results in discriminative behaviour of POMH to predict dysgraphia. We build a database with $w_\text{max}$ ranging from 23 to 39, $\hat{w}$ and the reconstruction error of the POMH model \texttt{dist} being dependent on $w_\text{max}$. In the end, by the classification algorithms, the optimal $w_\text{max}$ will be determined based on the prediction ability of the algorithms (Section~\ref{sec:classif}).

Knowing that the sampling is regular, we set the size of the "closing" morphological operator $w$ for a given symbol. For the model fit to be meaningful, we need a method for choosing $w$, the number and the instants of zero velocities for a symbol written by a child (Fig.~\ref{fig:density_nb_0_a}). 

%The corresponding number of zeros and three types of distance are created in the data base with $w_{\max}$ from 23 to 39, to study the effect of the choice of $w_{\max}$ on modeling and prediction.
\subsection{Number of zero velocity points: a critical parameter for the estimation of POMH}\label{sec:number of zero}

\begin{figure}[!ht]
    \centering
    \includegraphics[width =0.33\textwidth]{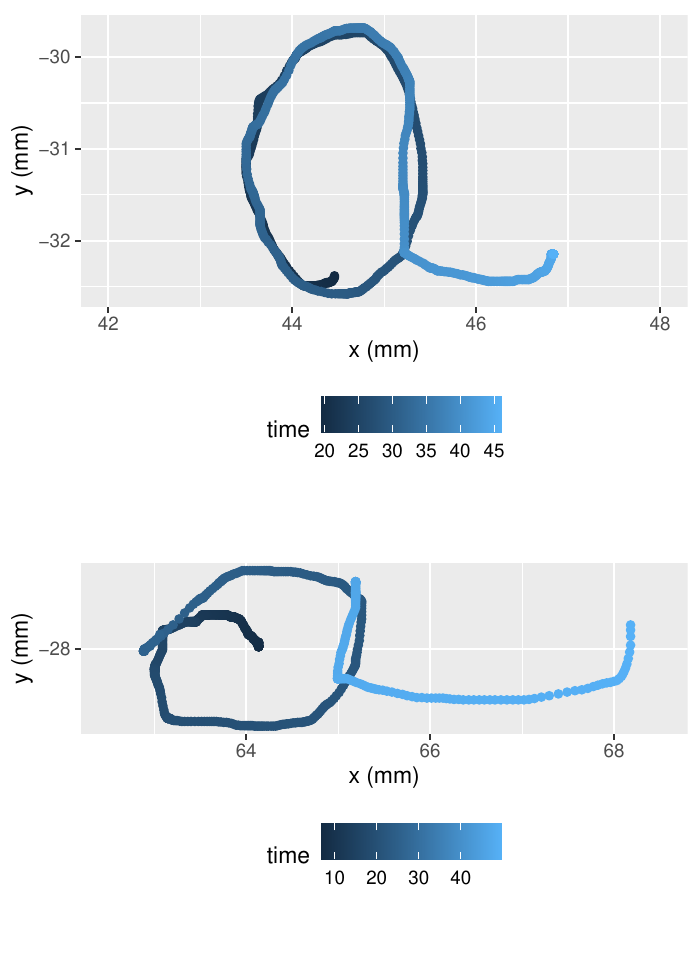}
    \includegraphics[width =0.66\textwidth]{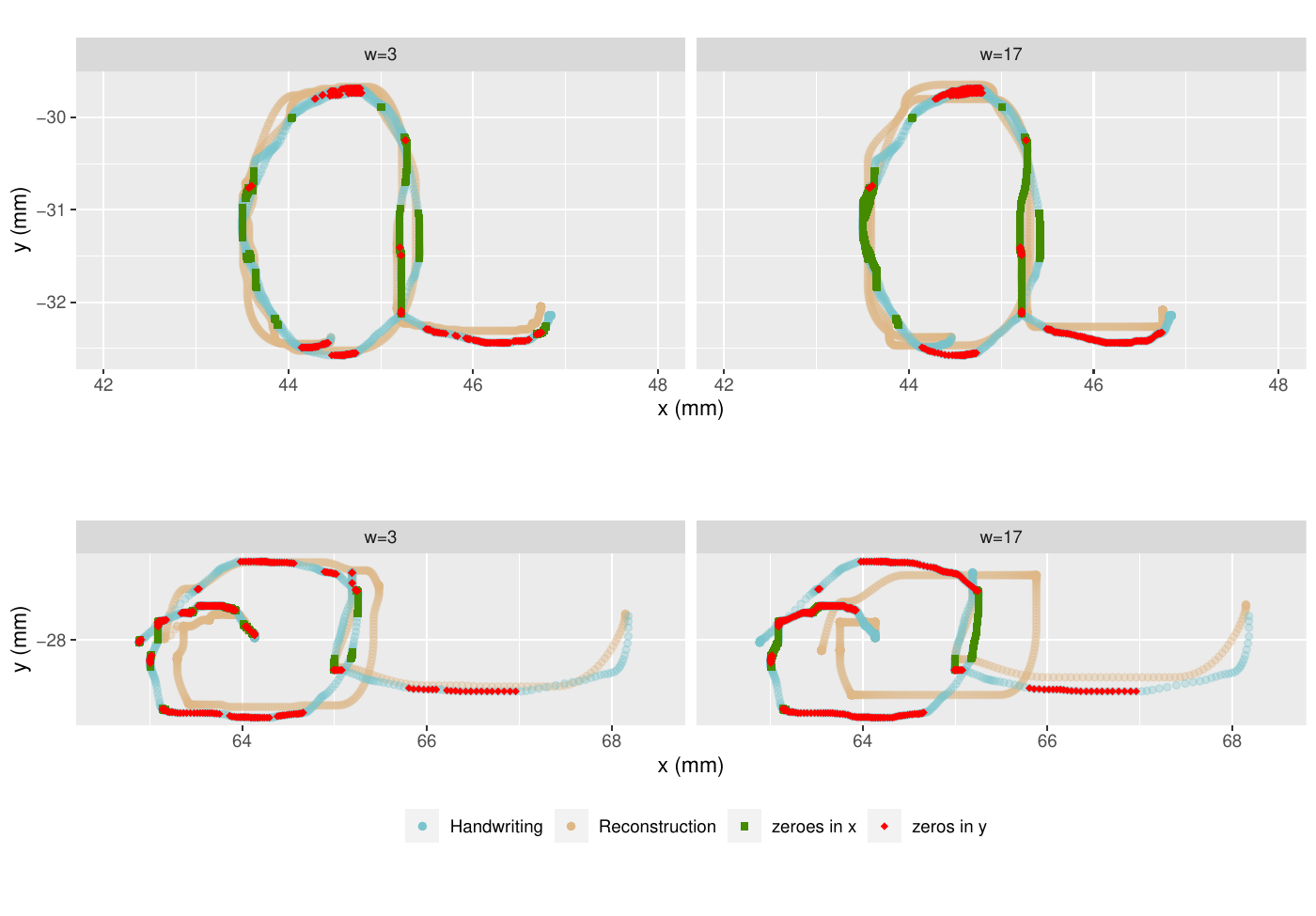}
    \caption{\textbf{Letter "a" written by a child with ID "H0217" without dysgraphia (top row) and by a child diagnosed with dysgraphia with ID "C0047" (bottom row).} Left: the recorded handwriting where color represents time; Middle (resp. Right): the original signal (light blue), the reconstructed signal (light orange) and moments of zero velocity in $x$ (green square) and in $y$ (red diamond) when the bandwidth of closing operator is $w=3$ (resp. $w=17$).}
    \label{fig:rec_many_less_zero}
\end{figure}

On the one hand, given that the temporal sampling rate of the tablet is stable ($\simeq200$Hz), the size $w$ of the structuring element of the closing operator has a direct impact on the number of extracted zero velocity moments, with larger sizes filtering out more zeros. For instance, the first row in Fig.~\ref{fig:rec_many_less_zero} corresponds to a symbol "\textit{a}" written by a child in second grade (CE1 in the French education system); with $w$ set to $3$ (equivalent to a time window of roughly 15ms), $20$ zeros are found in $x$ and $26$ in $y$, but with $w=17$ ($\simeq85ms$), the number of zeros respectively drops to $8$ and $9$. On the other hand, for each symbol, we expect a theoretical fixed number of zeros according to how handwriting is taught at school. However, more zeros do not necessarily mean that the child has dysgraphia, especially for young children who are still in the learning stage. Nevertheless, when children progress in the learning process, the number of zeros should decrease and converge to a fixed number for a given symbol (details in Section~\ref{sec:nbzero_vs_age}). This tendency is expected by handwriting professionals, even though at the end of the learning process, children usually personalize their handwriting and switch from only script to a mixture of cursive and script handwriting.

Obviously, the more zeros are included, the more break points in POMH, and the better the fit to the data. Yet, our method is based on the assumption that when imposing a low number of zeros for a given symbol, POMH can successfully reconstruct non-pathological handwriting but has more difficulty with children with dysgraphia. For instance, in Fig.~\ref{fig:rec_many_less_zero}, the panels on the left emphasize the writing process w.r.t time; the TD child on the top row wrote the letter "\textit{a}" with a single stroke (starting from the bottom, one and a half circle in clockwise direction before drawing the tail), while the child with dysgraphia (DG) on the bottom row wrote the letter in a more classical way (with two strokes). The other panels overlay the reconstructed trace (light orange) on top of the original trace (light blue), with zeros displayed for both $x$ (green) and $y$ (red). In the middle panels, with $w=3$, the numbers of zeros are $20$ in $x$, $26$ in $y$ for the first child, $23$ and $26$ for the second. With $w=17$ in the right panel, these numbers decrease to $8$ and $9$ for the first child, $7$ and $9$ for the second. Although the numbers of zeros are similar given the structuring element size, it can be noticed that the reconstruction with a lower number of zeros moves away drastically from the original trace for the child with dysgraphia (Fig.~\ref{fig:rec_many_less_zero} bottom right) and remains correctly estimated for the TD child (Fig.~\ref{fig:rec_many_less_zero} top right). The average distance (in millimeters) between the original and reconstructed traces increases from 0.1294 to 0.1509 ($+16\%$) for the TD child and from 0.2006 to 0.5006 ($+149\%$) for the child with dysgraphia.

When too less number of zeros is exploited, POMH will give bad reconstruction for both populations with and without dysgraphia. Therefore it is necessary to find a right number in the middle to make the reconstruction error discriminative for the two populations.

As it has been explained in the beginning of the section, the parameter $w$, size of the closing operator, has a direct impact on the number and positions of zeros. We have two ways of thinking about this problem. The first way is to look for the $w$ in order to have the optimal goodness of fit with the signal. Following this idea, we can use dynamical programming (see for example~\cite{bellman1961approximation}) to find the number of zeros and the instants of zero velocities. The second way, we look for the $w$ in order to have the "right" number of zeros, according to the canonical way of writing a symbol, giving that a child does not have dysgraphia. It is found that the "right" number of zeros for a given symbol is dependent on the age of a participant. It is also shown that children with dysgraphia have in average higher number of zeros. (Section~\ref{sec:nbzero_vs_age}) Therefore if we impose the same number of zeros for a child with dysgraphia as for a TD child of the same age, it will bias the reconstruction and make the error of reconstruction discriminative. In the following of this article, we explore the second way to estimate the parameter $w$.
%\jcq{Ci-dessus, j'ai introduit des tentatives de "normalisation" des termes pour les enfants (vu qu'en psychologie, surtout ici basé sur données, parler de normal vs. pathologique directement peut poser problème (même si ok côté JdS, autant s'entraîner pour l'article, et à confirmer/lisser par Caroline). Mais je ne l'ai pas fait partout... désolé. J'ai aussi tenté d'harmoniser traces/signals, stops/zeros}
%\jcq{Par habitude/principe, je préfère que les descriptions visuelles des figures soient dans la légende des figures, et que le texte décrivent les éléments quantifiés/chiffrés. Mais je n'ai pas réduit/modifié ici, vu que c'est une contrainte très dépendante des revues/conférences. Par contre, les "squares" et "crosses" en l'état, on ne les voient pas trop, sauf zoomer très fort (et encore, vu qu'images non-vectorielles ?).}
%\jcq{Toujours pour faciliter l'interprétation, j'ai rajouté "in mm" pour la distance, mais a-t-on confirmation que c'est bien ça l'unité, vu que je ne me souviens pas d'une confirmation explicite durant notre dernière réunion avec Caroline (oubli de ma part possiblement) ?}
%\jcq{On a répétition des résultats pour l'enfant non dys, mais pour que ça reste aussi clair, je ne vois pas comment réduire, donc ok pour moi comme ça.}

% "H0217" of a non child with dysgraphia in ce1 on the top line
% "C0047" for dysgrphafic child on the bottom line

\vspace{0.5cm}
In this section, we presented the conception of POMH, the problem of $w$ estimation arose in the application of POMH on real handwriting traces. In the next section, based on the assumption of the POMH model that the model can reconstruct handwriting traces if the movement is smooth, we aim to find the estimation of $w$ to make sure that the error of reconstruction is discriminative between populations with and without dysgraphia. 
\section{Transition from POMH to supervised classification}\label{sec:transi_pomh_classif}

We discuss in this section the creation of variables to include into the classification algorithm, i.e. the reconstruction error (the distance between the original trace and the one reconstructed by POMH model), the size of the closing operator $w$. As POMH decomposes the writing dynamics into two dimensions, there are two closing operators with size $w_x$ and $w_y$. To simplify the notation, we use $w$ to indicate both directions, unless there is a necessity to specify the dimensions.  
\subsection{Data description}
Children were asked to write cursively, without a time limit, the 26 letters of the Latin alphabet in cursive lower case, as well as the 10 digits, randomly dictated. The BHK test \cite{charles2004bhk} was also performed in order to identify children with dysgraphia. The dictation was performed on a digital Wacom\copyright Intuos 4 A5 USB graphic
tablet. A sheet of paper was placed on the tactile surface to create a familiar writing condition for the participants. The writing dynamics is recorded at a frequency of $200 Hz$ and a spatial resolution of $0.25 mm$. The handwritten production were then evaluated by psychometricians based on the BHK test. Among the 545 children, 479 ($88\%$) were typically developping children and 66 ($12\%$) were diagnosed as having dysgraphia. 

\begin{figure}
\centering
\includegraphics[width = 0.8 \textwidth]{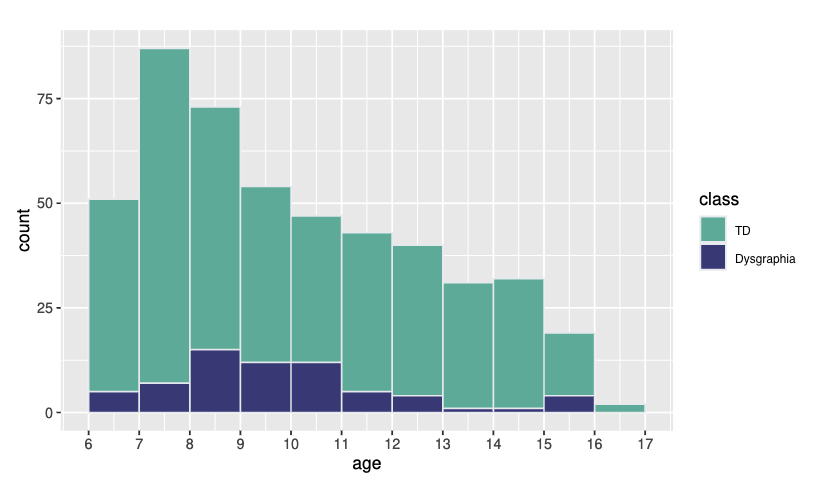}
\caption{The distribution of the age of participants in TD and dysgraphia groups.}
\label{fig:age_hist}
\end{figure}

The participants came from first grade to ninth grade, with age ranging from 6.5 to 16 years old. The distribution of the age of participants in the TD and dysgraphia groups is shown in Fig. \ref{fig:age_hist}.

Note that not all children have completed all the 36 symbols. Missing values occured when a participant didn't know how to write the requested symbol or when they wrote the wrong symbol, for example, a "5" instead of "2". The summary of missing rate in the two classes are given in Tables \ref{tab:NA_marginal} and \ref{tab:NA_bysymbol}.
\begin{table}
\centering
\caption{\label{tab:widgets1} \small The distribution of the non avaialble data (NA) in the two groups}
\label{tab:NA_marginal}
\small
\begin{tabular}{l|r|r}
&TD & Dysgraphia \\\hline
Complete&363 & 42 \\
At least one NA &116 & 24\\
Missing rate & 24.2\% & 36.4\%
\end{tabular}
\end{table}

\begin{table}
\centering
\caption{\label{tab:widgets2}The distribution of NA data for each symbol and each class.}
\label{tab:NA_bysymbol}
\scriptsize
\begin{tabular}{l|r|r|r|r|r|r|r|r|r|r|r|r|r|r|r|r}
Symbol & a & b & c & d & e & f & g & h & i & j & k & l & m & n & o & p  \\\hline
NAs TD & 1 & 0 & 2 & 7 & 2 & 3 & 14 & 17 & 3 & 9 & 39 & 3 & 2 & 3 & 0 & 1 \\
NAs Dys & 1 & 2 & 1 & 0 & 0 & 3 & 3 & 1 & 2 & 4 & 5 & 0 & 2 & 1 & 1 & 3 
\end{tabular}
\vspace{0.3cm}
\hspace{-0.5cm}
\begin{tabular}{r|r|r|r|r|r|r|r|r|r|r|r|r|r|r|r|r|r|r|r|r}
q & r & s & t & u & v & w & x & y & z & 0 & 1 & 2 & 3 & 4 & 5 & 6 & 7 & 8 & 9  \\\hline
43 & 4 & 2 & 2 & 3 & 5 & 48 & 23 & 39 & 18 & 1 & 1 & 2 & 1 & 2 & 2 & 1 & 2 & 2 & 1 \\
4 & 2 & 0 & 0 & 1 & 1 & 8 & 4 & 5 & 3 & 1 & 2 & 0 & 0 & 0 & 1 & 0 & 0 & 0 & 0 
\end{tabular}
\end{table}

\subsection{The number of zeros in $x$ and $y$ decreases when the age of participants increases}\label{sec:nbzero_vs_age}
%In this section, we aim to clarify the correlation between the number of zero velocity and the age of the participant. First of all, we sorted the participants in ascending order by age. To study the performance of children at age $x$, a window of six month is set, i.e., individuals with age $[x-3*30, x+3*30[$ are included.
%Secondly, as introduced in the beginning of Section 3, the number of zero velocity varies according to the bandwidth $w$ of the closing operator. It is observed that given a symbol, when the bandwidth $w$ increases, the statistics of the number of zero velocity (mean, quantiles) for a group of participants decrease. However, on the individual scale, there are usually plateau phenomenons. For instance, for individual $i$, when $w$ varies from $9$ to $15$, the number of zero velocity stays unchanged.

\begin{figure}
    \centering
    \includegraphics[width = \textwidth]{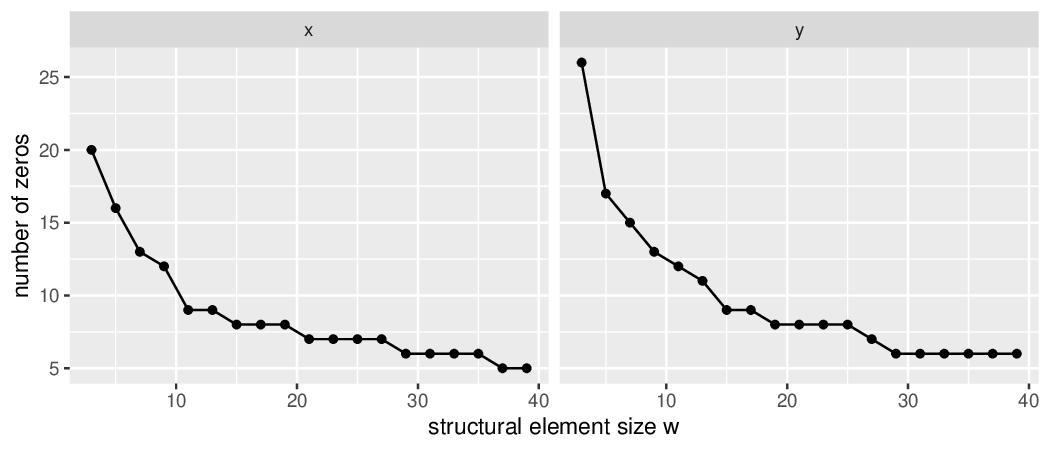}
    \caption{An illustration of the evolution of number of zero as a function of $w$ for symbol "a" by a TD child (ID "H0217")}
    \label{fig:nb0_w}
\end{figure}

Here we study the relationship between the age of the participants and the number of zeros, underlain by the problem of the estimation of $w$. For a given child and symbol, we expect the number of zeros to decrease when size $w$ increases but the acceleration decreases (Fig.~\ref{fig:nb0_w}). 

% \jcq{Légende à développer en effet, vu qu'impossible ici d'inférer ce qui est représenté : moyenne sur l'ensemble des symboles selon fenêtre ? Mais vu que fortement conditionné par l'âge vu figure 5, est-ce que c'est utile ? Vu la référence dans le texte, est-ce pour un seul enfant et symbole, et donc juste illustratif ? YJ: Yes one child one symbol for illustration purpose}
We study the number of zeros for each symbol for children at age $k \in [6.5,16]$ years (first to ninth grade, CP to troisième in the French educational system). To consider a sufficient amount of data, children with age $k\pm3$ months are considered. Therefore, if we set $w_\text{max}$, the maximum value that $w$ takes, and aggregate across children the number of zeros obtained for different values of $w$ in a wide enough range (in $\{3,5,\ldots,w_\text{max}\}$) for a given symbol and age range, numbers of zeros the best fitted to the raw signal should be over-represented in the resulting distribution.
For TD children, the vectors of number of zeros to be aggregated are
\begin{eqnarray*}
     &\left\{\left.\mathbf{n}^{(S)}_{x,i}(w_x)\right| a_i \in [a- 3 \text{ months}, a + 3 \text{ months}], w_x \in \left\{3, 5, \cdots, w_{max}\right\}\right\},&\\
     &\left\{\left.\mathbf{n}^{(S)}_{y,i}(w_y)\right| a_i \in [a- 3 \text{ months}, a + 3 \text{ months}], w_y \in \left\{3, 5, \cdots, w_{max}\right\}\right\},&\\
     %\mathbf{N}_{S,L,x}(a) = \texttt{median}(\mathbf{n}_i(a', w_x), a' \in [a- 3 \text{month}, a + 3 \text{month}]), w_x \in {3, 5, \cdots, L}), 
     %\mathbf{N}_{S,L,y}(a) = \texttt{median}(\mathbf{n}_i(a', w_y),  a' \in [a- 3 \text{month}, a + 3 \text{month}]), w_y \in {3, 5, \cdots, L})
\end{eqnarray*}

\noindent where $\mathbf{n}^{(S)}_{x,i}(w)$ (resp. $\mathbf{n}^{(S)}_{y,i}(w)$) is the number of zeros in $x$ (resp. $y$) dimension when a closing operator with size $w_x$ (resp. $w_y$) is applied on the signal of an individual $i$ and symbol $S$. For a given age $a$, children of age $a_i \in [a - 3 \text{ months}, a + 3\text{ months}]$ are included. For each individual, apply closing operator with $w_x$ (resp. $w_y$) from $3, 5, \text{to} \; w_\text{max}$. 
%Given that most of the dataset corresponds to TD children, this approach allows to parsimoniously extract the most representative ---and possibly meaningful--- number of zeros.

%\jcq{Range for the operator size to confirm (e.g. 1 = unfiltered not considered), and same for age range, here based on figure.}
%\jcq{J'ai retiré les détails sur l'individu "i", vu que non représenté/représentable sur les figures ?}
%\jcq{J'espère que l'explication et la "rationale" derrière l'agrégation brutale est suffisante. Pour la parcimonie, c'est relatif au besoin sinon d'introduire une autre étape / paramètres pour extraire la "bonne" largeur de fenêtre, mais c'est très discutable, et pas forcément clair en l'état.}

% Comment on peut bien expliquer le fait de inclure differents bandwidth dans le graphique ?
% JC: Oups, j'avais oublié ce point, mais je m'y attaque de suite (et j'arrêterai probablement là dessus, pour ce soir du moins)!
% Merci! Je reprends demain matin.
% JC: bonne soirée, je ne vais pas tarder non plus.
% Bonne soirée à toi auss (c'est rogolo)
% JC: fin du tchat Overleaf hors tchat Overleaf pour ce soir ;)

% Le nombre de zero 
% Comment ce modèle se comporte 

\begin{figure}[!ht]
    \centering
    \includegraphics[width=\textwidth]{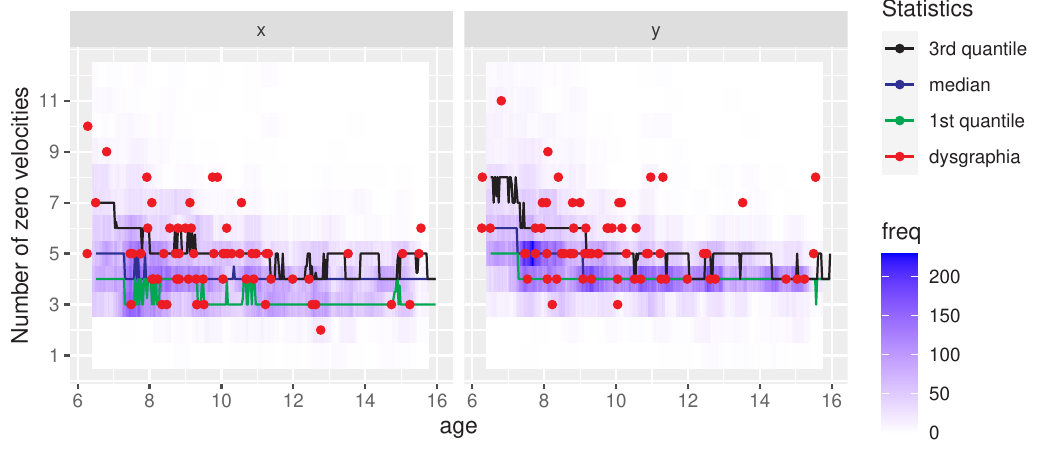}
    \caption{\textbf{Density and statistics of the number of zeros on $x$ and $y$-axis against age, for letter "a".} For TD children, density (blue shade), $Q1$ (green), median (blue line) and $Q3$ (black) of the number of zeros in $x$ (left) and in $y$ (right), aggregated across values of $w$; For individuals with dysgraphia, median (red points).
    %\jcq{Je pensais que pour les zéros des dysgraphiques, on pouvait les superposer sur le graphique du haut, mais vu que c'est pour des $w$ qui varient, ça serait en effet illisible, et l'approche retenue semble donc la bonne. A confirmer si on peut faire mieux pour article ultérieur, mais ça semble très bien pour l'instant ici (même si ça prend un peu de place en plus, mais on n'est plus à a près ;).}
    }
    \label{fig:density_nb_0_a}
\end{figure}

%\jcq{Fin de relecture pour moi pour aujourd'hui, et je suis en réunion à l'IMAG sur un autre projet toute la journée de demain (mardi) :( Ne m'attendez donc pas pour transmettre pour relecture aux collègues. NB: si vous voulez virez tous mes commentaires, il suffit de (dé)commenter la commande tout en haut du TeX.}
    
%If we take an example of the symbol "a" (Fig.~\ref{fig:density_nb_0_a}), the number of zero velocity at different quantiles (green: the first quantile, blue: the median and black: the third quantile) decreases as age of the participant increases. The variance also decreases with age, as the blue shade band is wide at low ages and becomes narrow as age increases. 
Note that only the TD population is included for the study of the relation between number of zeros and age. If we take the symbol "a" as an example (Fig.~\ref{fig:density_nb_0_a}), $\mathbf{n}^{(S)}_{x,i}(w)$ and $\mathbf{n}^{(S)}_{y,i}(w)$ are represented by blue shade, the three quartiles for the number of zeros decreases as the age of participants increases. Dispersion also decreases with age, as reflected by the density.
%Each red point in Fig.~\ref{fig:density_nb_0_a} represents, for un individual with dysgraphia, the median of the numbers of zero velocity w.r.t. different $w$s when he writes the letter "a". Compared to the TD children, it can be observed that there are more red points above the blue line (the median for TD children) than below. This means that children suffering from dysgraphia stop more often during writing than the average level of a TD children of the same age.

Each red point represents the median of the numbers of zeros (aggregated across sizes~$w$) for one child with dysgraphia, when writing letter "a". 
\begin{eqnarray*}
        &\mathbf{N}^{(S)}_x(i) = \texttt{median}\left(\left\{\left.\mathbf{n}^{(S)}_{x}(w_x, i)\right|w_x \in \{3, 5, \cdots, w_{max}\}\right\}\right),\\
     &\mathbf{N}^{(S)}_y(i) = \texttt{median}\left(\left\{\left.\mathbf{n}^{(S)}_{y}(w_y, i)\right|w_y \in \{3, 5, \cdots, w_{max}\}\right\}\right)&
\end{eqnarray*}
where $i$ indicates individual $i$ with dysgraphia.
Compared to TD children, it can be observed that there are more red points above the blue line (median for TD children) than below. The associated children therefore stop or oscillate more than the average TD child of the same age.

The value of $w_\text{max}$ is chosen empirically. To find the optimal value, we created a database with $w_\text{max}$ ranges from 23 to 39 and through cross validation and select the value which gives the maximum prediction accuracy of dysgraphia by the classification algorithms (Section~\ref{sec:classif}). 

\subsection{Estimate $\hat{w}$ with the number of zeros fixed for a given age and symbol}\label{subsec:estimate_w}

Besides using construction error (\texttt{dist}) as discriminative variable, one should find a fair way to choose $w_x$ and $w_y$. The values of  $w_x$ and $w_y$ have an impact on the construction error, as the values decide the number and the positions of zeros in the velocity signal. If the values are too small, then the construction error is small for both dysgraphic writing traces and non-dysgraphic ones. The variable \texttt{dist} becomes therefore non-discriminative. Same for the case where the values of $w_x$ and $w_y$ is too big, then the construction error \texttt{dist} are too big for both dysgraphic writing traces and non-dysgraphic ones. It is therefore necessary to find values of $w_x$ and $w_y$ to make variable \texttt{dist} discriminative.

In addition, according to Fig.~\ref{fig:density_nb_0_a}, due to the natural learning process, for a given letter, the number of zeros decreases with age. Therefore we also want to take into account the effect of age when estimating $w_x$ and $w_y$.

With TD children, we construct a database of the number of zeros $\mathbf{N}^{(S)}(a)$ as a function of age $a$ for each symbol $S$.
\begin{eqnarray*}
     &\mathbf{N}^{(S)}_x(a) = \texttt{median}\left(\left\{\left.\mathbf{n}^{(S)}_{x,i}(w_x)\right| a_i \in [a- 3 \text{ months}, a + 3 \text{ months}], w_x \in \{3, 5, \cdots, w_{max}\}\right\}\right),&\\
     &\mathbf{N}^{(S)}_y(a) = \texttt{median}\left(\left\{\left.\mathbf{n}^{(S)}_{y,i}(w_y)\right| a_i \in [a- 3 \text{ months}, a + 3 \text{ months}], w_y \in \{3, 5, \cdots, w_\text{max}\}\right\}\right)&
     %\mathbf{N}_{S,L,x}(a) = \texttt{median}(\mathbf{n}_i(a', w_x), a' \in [a- 3 \text{month}, a + 3 \text{month}]), w_x \in {3, 5, \cdots, L}),  \\
     %\mathbf{N}_{S,L,y}(a) = \texttt{median}(\mathbf{n}_i(a', w_y),  a' \in [a- 3 \text{month}, a + 3 \text{month}]), w_y \in {3, 5, \cdots, L})
\end{eqnarray*}
 then the number of zeros on $x$ (resp. on $y$) for age $a$ and for a given symbol $S$, is the median over all. It can be noticed that the number of zeros $\mathbf{N}^{(S)}_{x}(a)$ depends on the symbol $S$ and age $a$.

%The way to fix $\mathbf{N}^{(S)}_x(a)$ seems a bit arbitrary. The reason why it should work is that $\mathbf{n}^{(S)}_{x,i}(w_x)$ decrease in the form of stairs as $w_x$ increases (Fig.~\ref{fig:nb0_w}). For instance, as the operator $w_x=3$ is applied, $\mathbf{n}^{(S)}_{x,i}(w_x)$ decreases compared to the raw signal (with no operator applied), let's say the next change point is $w_x=11$, then $\mathbf{n}^{(S)}_{x,i}(w_x)$ keeps unchanged as $w_x$ increases from $3$ to $9$, and if $w_x$ continues to increase, then $\mathbf{n}^{(S)}_{x,i}(w_x)$ decrease and then keep unchanged for some $w_x$ from $11$ to the next change point. Around the "right" value of $\mathbf{N}^{(S)}_x(a)$, it should keep unchanged for a large scale of $w_x$. Therefore we believe that by fixing the "right value" of $w_\text{max}$, we can find $\mathbf{N}^{(S)}_x(a)$ by calculating the median over all $w_x \in \{3, 5, \cdots, w_\text{max}\}$. It then becomes obvious that the parameter $w_\text{max}$ is crucial for finding $\mathbf{N}^{(S)}_x(a)$. In Section~\ref{sec:classif}, we test the classification algorithms on a wide range of $w_\text{max}$, from $23$ to $39$ and choose the optimal $w_\text{max}$ which gives the best prediction results.

Once the reference database $\mathbf{N}^{(S)}_{x}(a)$ is built, the estimation of $w$ for individual $i$
\begin{equation*}
\widehat{w}^{(S)}_{x,i} = \min_{w_x}\left[\mathbf{n}^{(S)}_{x,i}(w_x)\leq \mathbf{N}^{(S)}_{x}(a_i)\right],
\end{equation*}
which says that we fix the number of zeros on both $x$ and $y$ dimensions according to age and find the minimum $w_x$ and $w_y$ which allow to obtain these number of zeros. 

\subsection{Different types of distance}\label{subsec:ty_dist}

There are different ways to calculate the distance between the original handwriting trace $\left(x_t,y_t\right)_{1\leq t\leq T_{i,S}}$ and the reconstructed one $\left(\hat{x}_t,\hat{y}_t\right)_{1\leq t\leq T_{i,S}}$ with $T_{i,S}$ the number of points used by the individual $i$ for the symbol $S$. Three different types of distance are defined below. In the next section, by cross validation in classification algorithms, the most discriminatory type of distance will be selected.

\noindent The $\mathcal{L}_1$ distance:
\begin{equation*}
    \frac{1}{T_{i,S}}\sum_{t=1}^{T_{i,S}}\left(\left|x_t-\hat{x}_t\right| + \left|y_t - \hat{y}_t\right|\right).
\end{equation*}
The $\mathcal{L}_2$ distance:
\begin{equation*}
    \frac{1}{T_{i,S}}\sum_{t=1}^{T_{i,S}}\sqrt{\left(x_t-\hat{x}_t\right)^2 + \left(y_t - \hat{y}_t\right)^2}.
\end{equation*}
The $\mathcal{L}_{\infty}$ distance:
\begin{equation*}
    \frac{1}{T_{i,S}}\sum_{t=1}^{T_{i,S}}\left[\max\left(\left|x_t-\hat{x}_t\right|, \left|y_t - \hat{y}_t\right|\right)\right].
\end{equation*}

To summarise, by applying the POMH model on handwriting traces, we extract the construction error (noted \texttt{dist}), $\widehat{w}_x$, $\widehat{w}_y$ as features for the classification algorithms. There are two other parameters to be determined by the classification algorithms, which are the type of distance and $w_\text{max}$.

\section{two-layer statistical modeling}\label{sec:classif}
%\todo[nolist, inline, color = purple]{Ajoute un schéma je pense. Notamment, comment tu divises ta base à ce moment là pour bien montrer comment les données sont utilisées. Cela doit aller de pair avec le paragraphe sur les données.}

Even though more complicated writing trace data are available (sentences or paragraphs), we limit the present analysis to individual symbols (26 letters + 10 digits), which are easier to model and still can reveal interesting and fundamental information/rules about handwriting and its learning process. 

\subsection{Disadvantages of the one layer model}

At the very beginning of this study, trials have been carried out to detect dysgraphia with one model. From each symbol we extract \texttt{dist}, $w_x$ and \texttt{total$\_$time}. For a participant having written the 36 symbols, the inputs of the model are three variables for all 36 symbols and the grade of the participant (109 variables in total, shown in Tab. \ref{data_onelayermodel}). The output of the model is a binary variable $Y_i$ indicating whether the participant is classified as having dysgraphia or not. Because of the large number of variables compared to the number of observations ($\sim$500), and collinearity among variables, it is difficult to interpret the model and the prediction capacity was not satisfying.

\begin{table}[ht]
\tiny
\centering
\caption{Dataset for a one-layer model} 
\begin{tabular}{c|ccc|ccc|ccc|ccc|c|c}
    \multirow{2}{*}{ID}&\multicolumn{3}{c|}{a}&\multicolumn{3}{c|}{b}&\multicolumn{3}{c|}{$\cdots$}&\multicolumn{3}{c|}{9}&\multirow{2}{*}{\texttt{section}}&\multirow{2}{*}{\texttt{dys}}\\
    \cline{2-13}
         & $w_x$ &\texttt{dist} & \text{tt} & $w_x$ &\texttt{dist}& \texttt{tt} & $w_x$ & \texttt{dist} & \texttt{tt}& $w_x$ &\texttt{dist} & \texttt{tt} & &\\ \hline 1&$X^{(a)}_{1,1}$&$X^{(a)}_{1,2}$&$X^{(a)}_{1,3}$&$X^{(b)}_{1,1}$&$X^{(b)}_{1,2}$&$X^{(b)}_{1,3}$&\multicolumn{3}{c|}{$\cdots$}&$X^{(9)}_{1,1}$&$X^{(9)}_{1,2}$&$X^{(9)}_{1,3}$&$X_{1,4}$&$Y_1$\\
        2&$X^{(a)}_{2,1}$&$X^{(a)}_{2,2}$&$X^{(a)}_{2,3}$&$X^{(b)}_{2,1}$&$X^{(b)}_{2,2}$&$X^{(b)}_{2,3}$&\multicolumn{3}{c|}{$\cdots$}&$X^{(9)}_{2,1}$&$X^{(9)}_{2,2}$&$X^{(9)}_{2,3}$&$X_{2,4}$&$Y_2$\\
        $\vdots$&\multicolumn{3}{c|}{$\vdots$}&\multicolumn{3}{c|}{$\vdots$}&\multicolumn{3}{c|}{}&\multicolumn{3}{c|}{$\vdots$}&$\vdots$&$\vdots$\\
        n&$X^{(a)}_{n,1}$&$X^{(a)}_{n,2}$&$X^{(a)}_{n,3}$&$X^{(b)}_{n,1}$&$X^{(b)}_{n,2}$&$X^{(b)}_{n,3}$&\multicolumn{3}{c|}{$\cdots$}&$X^{(9)}_{n,1}$&$X^{(9)}_{n,2}$&$X^{(9)}_{n,3}$&$X_{n,4}$&$Y_n$ 
\end{tabular}
\label{data_onelayermodel}
\end{table}

Finally, we adopted a model with two layers: the first layer contains 36 classifiers, one for each symbol. The inputs of one classifier are: \texttt{dist}, \texttt{section}, $\widehat{w}_x$, \texttt{total$\_$time} and the output is the probability that the symbol was written by an individual with dysgraphia. The second layer is the modeling in the individual level and determines whether one participant is classified as having dysgraphia, using as input the probability of dysgraphia for all the symbols written by the participant.

In section~\ref{subsec:classifier}, we present the principle of the algorithms used before explaining the structure of our two-layer modeling in section~\ref{subsec:composition}.

\subsection{Classification algorithms}\label{subsec:classifier}

Here, we present the four classification algorithms used in the procedure. Notations are valid only in the section of each model description and should not be generalized to the rest of the article.

\paragraph{Generalized Linear Model (GLM)}

The Generalized Linear Models (GLM) extend the linear models to the case where the response variable is not normally distributed (see for example~\cite{nelder1972generalized}). Given a binary results $Y$, for example having dysgraphia (1) or not (0), and co-variables $x_1,\ldots,x_p$, the model assumes that the observation $Y$ is the result of a binary law $\mathcal{B}\left(\text{logistic}(\eta)\right)$ where 
\begin{equation}\label{eq:GLM}
    \text{logistic}(\eta)= \frac{e^\eta}{1+e^\eta}\text{ and }\eta = \theta_0 + \theta_1 x_1 + \cdots + \theta_p x_p.\\
\end{equation}

Thus, the greater the $\eta$ value, the greater the probability that $Y$ is worth one. Parameters $\btheta=\left(\theta_j\right)_{0\leq j\leq p}$ of Equation~\eqref{eq:GLM} are estimated by maximizing likelihood using a Newton Raphson-type iterative algorithm. Since we are in a prediction problem, we associate parameter estimation with variable selection using the Akaike Information Criterion or criterion AIC (see \cite{zhang2016variable}).

At the end, the procedure returns a parameter estimate (with zero values for unselected variables) which, given co-variables $x_1,\ldots,x_p$, estimates the probability that the associated variable $Y$ is worth 1.

\paragraph{Random Forest (RF)}
Random Forest is an ensemble method that contains a certain number of Decision Trees (see for example~\cite{breiman1996bagging}). An ensemble method, to make it simple, is by combining a bunch of weak learners to make a strong learner. A weak learner here is a Decision Tree. To be noticed that Decision Trees are not grown on the entire sample and not all predictors are used for each node of each tree, but on sub-samples using randomly sampled predictors. About two third of observations are selected by bootstrap sampling with replacement and the one third left is known as Out-of-Bag (OOB) observations. In classification problems, the number of sampled variables as candidates for splits is $\sqrt{p}$ where $p$ is the number of total variables (see~\cite{genuer2008random}). An the end, $N$ Decision Trees are trained on random samples and the error calculated on OOB observations. Every Decision Tree is a vote and Random Forest returns for an individual $i$ the class with the highest vote when individual $i$ is in OOB samples. Consequently, the probability $\mathbb{P}(Y_i = 1)$ is the proportion of trees predicting $Y_i = 1$ when the individual $i$ is in the OOB sample. 

\paragraph{Support Vector Machine (SVM)}SVM is a machine learning algorithm aiming at finding a linear hyperplane to separate two populations with maximum margin \cite{boser1992training}. The points closest to the hyperplane are called Support Vectors and the distance from Support Vectors to the hyperplane is defined as margin.

In the case where the problem is linearly separable, which means there exists a hyperplane to perfectly separate two populations without errors, the Support Vector Machine is the hyperplane that separate two populations with the maximum margin, to ensure that it will generalize well on other datasets.

However, in real life, it is rare to find a problem that is linearly separable. Therefore, the notion of soft margin was introduced. The problem becomes finding a hyperplane which maximizes the margin and minimizes the penalty of wrong classification defined as $C$ mutiplied by the distance of the example to the other side of the margin. The parameter $C$ therefore controls the trade-off between the wrong classification penalty and the size of the margin: When $C$ is small, more mis-classifications are allowed and the margin is big; on the contrary, when $C$ is big, less mis-classifications are allowed but the margin is smaller.

Another powerful feature about SVM is its Kernel trick. When data is not linearly separable, SVM allows to project data into feature space, usually higher dimension, using Kernel trick. In the the feature space, it is easier to find a hyperplane to separate data.

The common non-linear kernels are Polynomial kernel, Sigmoidal Kernel, and Radial Basis Function Kernel. In our application, the RBF Kernel is used due to its versatility and it has fewer parameters to tune. The RBF Kernel is given by

\begin{equation*}
   K(x^i, x^j) = \exp{(-\gamma\|x^i-x^j\|^2)},
\end{equation*}
where $x^i$ and $x^j$ are two samples from input space. The parameter $\gamma$ determines the scale of the influence that each support vector has on the decision boundary. A small $\gamma$ value means a wider RBF kernel and leads to a softer decision boundary with a larger margin, which allows for more mis-classifications. Conversely, a large $\gamma$ value leads to a more complex and localized decision boundary that fits the training data more closely, potentially resulting in overfitting.

It has been shown that the calculation time for SVM is closely related to the number of support vectors and increases rapidly as the sample size increase. The method is adapted to the middle size sample of around 500 observations that we have.

\paragraph{Counting }
The first layer returns the probability that each symbol is written by someone with dysgraphia $\widehat{p}^{(S)}_{i}, S = {a, b, \cdots, 9}, i = 1, 2, \cdots, n_S$. We set a threshold on the probability  $\widehat{t}^{(S)}_{\alpha} = Q_{\alpha}(\widehat{p}^{(S)}_{i}), i = 1, 2, \cdots, n_S$, which gives
\begin{equation}
\left\{   
\begin{array}{ll}
      Y^{(S)}_{i} = 1 & \text{if } \widehat{p}^{(S)}_{i} \geq \widehat{t}^{(S)}_{\alpha}, \\
        Y^{(S)}_{i} = 0 & \text{else},
\end{array}\right.
\label{eq:count_alpha}
\end{equation}
where $Q_{\alpha}(\cdot)$ is the $\alpha$ order empirical quantile function and $\widehat{t}^{(S)}_{\alpha}$ is the $\alpha$ order probability threshold. 
% \textbf{\textcolor{red}{Warning on $\widehat{p}^{(S)}_{\alpha}$ since index used for both $i$ and $\alpha$ of different nature/meanings, and not sure about the estimate everywhere?}}
% \textbf{\textcolor{blue}{Yunjiao: I then change $\widehat{p}^{(S)}_{\alpha}$ to $\widehat{t}^{(S)}_{\alpha}$. Both $\widehat{t}^{(S)}_{\alpha}$ and $\widehat{p}^{(S)}_{i}$ are estimations as they base on the output of the first layer.}}

For a given $\alpha$, we obtain, for each individual, the number of symbols considered as written by someone with dysgraphia (called positive symbols for simplicity). Then if the number of positive symbols is above the cut-off value, then the individual is considered as having dysgraphia.

\subsection{two-layer model}\label{subsec:composition}

We detail here our proposed two-layer procedure.

\paragraph{The first layer}

%We fix a symbol (for example the letter "a") and for each children $i$, we estimate the distance between the reconstruction and the data knowing an operator $w$. In this case, we obtain the following matrix:
The principle of the first layer is to look at the features for each symbol $S\in\{a,b,\ldots,z,0,1,\ldots,9\}$ that would predict if it was written by a child with dysgraphia. Thus, the features of each symbol are extracted from Table~\ref{data_onelayermodel} to keep only the operator size estimate $\widehat{w}_x$, \texttt{dist}, \texttt{section}, total time\footnote{\texttt{total$\_$time}: the time taken to write the symbol, time of pen in the air excluded} and the binary response variable, whether the individual who wrote the symbol has dysgraphia or not (see Table~\ref{tab:data_secondlayermodel_bysymbol}). Note that $X^{(S)}_{i,1} \in \{3, 5, \cdots, w_\text{max}\}$  are discrete quantitative, $X_{i,4} \in \{$1st, 2nd, 3rd, 4th, 5th, 6th, 7th, 8th, 9th-grade$\}$ qualitative ordinal and  $X^{(S)}_{i,2} $ and $X^{(S)}_{i,3}$ are continuous variables. In the first layer, the classification methods GLM, RF and SVM are tested (see Methods in Section~\ref{subsec:classifier} and Results in Section~\ref{sec:res}).\\

\begin{table}[!ht]
\centering
\caption{Data structure for a symbol $S\in\{a,b,\ldots,z,0,1,\ldots,9\}$. $n_S$ represents the number of individuals who have written the symbol. \label{tab:data_secondlayermodel_bysymbol}} 
    $\begin{array}{c|cccc|c}
        \text{ID children}&\text{operator size $\hat{w}_x$}&\text{distance}& \text{total time}&\text{section} &\text{dysgraphia}\\\hline
        1&X^{(S)}_{1,1} \coloneqq \widehat{w}_{x,1}&X^{(S)}_{1,2}&X^{(S)}_{1,3}&X_{1,4}&Y_1\\
        2&X^{(S)}_{2,1} \coloneqq\widehat{w}_{x,2}&X^{(S)}_{2,2}&X^{(S)}_{2,3}&X_{2,4}&Y_2\\
        \vdots&\vdots&\vdots&\vdots&\vdots&\vdots\\
        n_S&X^{(S)}_{n_S,1} \coloneqq \widehat{w}_{x,n_S}&X^{(S)}_{n_S,2}&X^{(S)}_{n_S,3}&X_{n_S,4}&Y_{n_S}\\
    \end{array}$  
\end{table}

\paragraph{The second layer}

The second layer is about the detection of dysgraphia at the individual level, using as input the classification results of all single symbols issued from the first layer. 

Three methods are used here in the second layer, including Counting, GLM or RF (see Section~\ref{subsec:classifier}). Counting is the most intuitive one, in which it is assumed that all symbols have the same distinguish ability, on the contrary, the other two classifiers assume different distinguish ability of symbols. SVM is not used in the second layer and the reason is explained later in the end of Section~\ref{subsec:svm+counting}

% Following the second layer, we calculate the area under the ROC (Receiver Operating Characteristic) curve. 

%  and determine parameters $w_\text{max}$ and \texttt{ty}$\_$\texttt{dist} as the ones which maximize the AUC value

When using GLM or RF in the second layer, missing values (individuals who didn't write all symbols) are excluded. The distribution of NAs in the two classes and by symbol (Tables~\ref{tab:NA_marginal} and~\ref{tab:NA_bysymbol}) shows that the missing rate is higher in the Dysgraphia group than in the Non-dysgraphia one. The output of the second layer is the probability $\widehat{P}_i$ that each individual has dysgraphia. The AUC value is calculated by applying the cut-off on $\widehat{P}_i$: 
\[\left\{   \begin{array}{ll}
      \widehat{Y}_i = 1 & \text{if }\widehat{P}_i > \text{cutoff}, \\
         \widehat{Y}_i = 0 & \text{else}.
    \end{array}\right.\]
When the Counting method is used, the individuals who didn't write all symbols can be kept, which is an advantage of the method. Two assumptions can be made on missing values, the first one is that the symbols are missing at random; the second one is that participants tend not to write the symbol when they don't know or are not sure how to write it. The proportion of positive symbols ($n_{i,p}/n_i$, where $n_{i,p}$ is the number of positive symbols written by individual $i$ and ${n_i}$ is the total number of symbols written by individual $i$) is calculated to determine whether the individual $i$ has dysgraphia or not. We then apply the cut-off value on the proportion of positive symbols $n_{i,p}/n_i$,
\[\left\{   \begin{array}{ll}
      \widehat{Y}_i = 1 & \text{if }n_{i,p}/n_i > \text{cutoff}, \\
         \widehat{Y}_i = 0 & \text{else}.
    \end{array}\right.\]
By varing the cutoff value from 0 to 1, we obtain the ROC curve and then accordingly the AUC value.

%Under the first assumption, the cut-off value should be based on the proportion of positive symbols ($n_{i,p}/n_i$, where $n_{i,p}$ is the number of positive symbols written by individual $i$ and ${n_i}$ is the total number of symbols written by individual $i$). Under the second assumption, we will argue that the optimal way is to apply the cut-off value also on the proportion of positive symbols. If the cut-off value is applied on the number of positive symbols, then it means that the missing values are counted systematically negative, which is counter-assumption. If someone wrote one symbol with hesitation, then it is more probably that the symbol classified as positive by the algorithm. If we count the missing values systematically positive, then the threshold should be applied on $(n_{i,p} + n_{i,m})/36$ where is the number of missing symbols by individual $i$. Therefore using $n_{i,p}/n_i$ or $(n_{i,p} + n_{i,m})/36$ is more adapted under the second assumption. 

We perform a comparison of methods. Apart from building the model which maximizes the AUC value, by training the model, we aim at finding the optimal parameter $w_\text{max}$ and we determine which of the three types of distance can the best distinguish between children with or without dysgraphia. For each combination, we obtain the AUC values as a function of $w_\text{max}$ and of the types of distance $\texttt{ty}\_\texttt{dist}$. We discuss also the way to find the optimal balance between TPR (True Positive Rate) and TNR (True Negative Rate) in Appendix~\ref{app:fpr_tpr} in the Supplementary Materials.

%and finally, the optimal results (AUC, FNR, FPR, ACC) with optimal parameters like in Fig.~\ref{fig:auc_acc}.

The structure of the two-layer classifier and the classification algorithms used in each layer are presented. In the next section, the classification results of five classifiers are compared, which are GLM then Counting, GLM then GLM, RF then Counting, RF then RF, SVM then Counting (see table~\ref{tab:summarise:layers} for a summary).

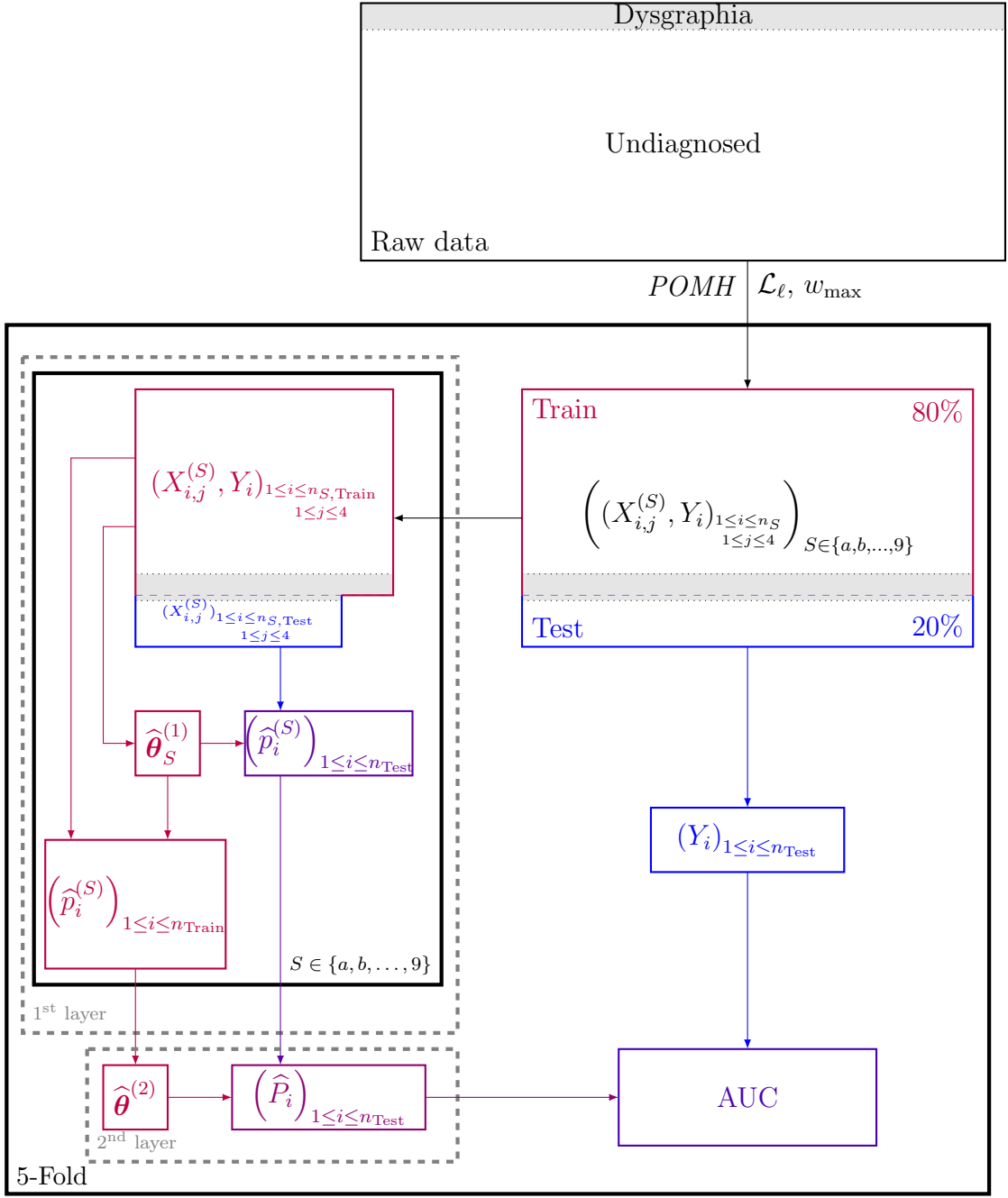
\begin{figure}[tp]
\vspace*{-4.5cm}
    \centering
    \begin{tikzpicture}[scale=0.95]

%%% Initial dataset
\draw (-2.5,4) node[above right] {Raw data};
% Separation
\fill[gray,opacity=0.2] (-2.5,7.585185) rectangle (7.5,8);
\draw[dotted] (-2.5,7.585185) -- (7.5,7.585185);
\draw (2.5,7.792593) node{Dysgraphia};
\draw (2.5,5.792593) node{Undiagnosed};
% Rectangle
\draw[thick] (-2.5,4) rectangle (7.5,8);

%%% Transformation
% Separation en deux
\fill[gray,opacity=0.2] (0,-0.868148) rectangle (7,-1.282963);
\draw[dotted] (0,-0.868148) -- (7,-0.868148);
\draw[dotted] (0,-1.282963) -- (7,-1.282963);
\draw (3.5,0) node{$\left((X^{(S)}_{i,j},Y_i)_{\overset{1\leq i\leq n_S}{\underset{1\leq j\leq 4}{}}}\right)_{S\in\{a,b,\ldots,9\}}$};
% Rectangle
\draw[purple,thick] (0,-1.2) -- (0,2) -- (7,2) -- (7,-1.2);
\draw[blue,thick] (0,-1.2) -- (0,-2) -- (7,-2) -- (7,-1.2);
\draw[dashed,purple!50!blue,opacity=0.5] (0,-1.2) -- (7,-1.2); %%% Essayer de rendre plus visible
\draw[purple] (7,2) node[below left] {80\%};
\draw[purple] (0,2) node[below right] {Train};
\draw[blue] (7,-2) node[above left] {20\%};
\draw[blue] (0,-2) node[above right] {Test};
\draw[->,>=latex] (3.5,4) -- (3.5,2) node[pos=0.2,left]{\textit{POMH}} node[pos=0.2,right]{$\mathcal{L}_{\ell},\,w_{\max}$};
\draw[->,>=latex] (0,0) -- (-2,0);

%%% Transformation
% Separation en deux
\fill[gray,opacity=0.2] (-6,-1.282963) -- (-2.8,-1.282963)-- (-2.8,-1.2)-- (-2,-1.2) -- (-2,-0.868148) -- (-6,-0.868148)  -- cycle;
\draw[dotted] (-6,-0.868148) -- (-2,-0.868148);
\draw[dotted] (-6,-1.282963) -- (-2.8,-1.282963)-- (-2.8,-1.2)-- (-2,-1.2);
\draw[purple] (-4,0.4) node{$(X^{(S)}_{i,j},Y_i)_{\overset{1\leq i\leq n_{S,\text{Train}}}{\underset{1\leq j\leq 4}{}}}$};
% Rectangle
\draw[purple,thick] (-6,-1.2) -- (-6,2) -- (-2,2) -- (-2,-1.2) -- (-2.8,-1.2);
\draw[blue,thick] (-6,-1.2) -- (-6,-2) -- (-2.8,-2) -- (-2.8,-1.2);
\draw[dashed,purple!50!blue,opacity=0.5] (-6,-1.2) -- (-2,-1.2); %%% Essayer de rendre plus visible
\draw[blue] (-4.4,-1.6) node{\tiny $(X^{(S)}_{i,j})_{\overset{1\leq i\leq n_{S,\text{Test}}}{\underset{1\leq j\leq 4}{}}}$};

%%% Reference
\draw[blue] (3.5,-5) node{$\left(Y_i\right)_{1\leq i\leq n_\text{Test}}$};
\draw[blue,thick] (2,-4.5) rectangle (5,-5.5);
\draw[blue,->,>=latex] (3.5,-2) -- (3.5,-4.5) ;

%%% Couche 1
% Theta 1
\draw[purple] (-5.5,-3.5) node{$\widehat{\boldsymbol{\theta}}^{(1)}_{S}$};
\draw[purple,thick] (-6,-3) rectangle (-5,-4);
\draw[purple,->,>=latex] (-6,-0.1333333) -- (-6.5,-0.1333333) -- (-6.5,-3.5) -- (-6,-3.5) ;
% ps train
\draw[purple,thick] (-7.4,-5) rectangle (-4.6,-7);
\draw[purple] (-6,-6) node{$\left(\widehat{p}^{(S)}_{i}\right)_{1\leq i\leq n_{\text{Train}}}$};
\draw[purple,->,>=latex] (-6,0.9333333) -- (-7,0.9333333) -- (-7,-5) ;
\draw[purple,->,>=latex] (-5.5,-4) -- (-5.5,-5);
% ps test
\draw[purple!50!blue,thick] (-4.3,-3) rectangle (-1.7,-4);
\draw[purple!50!blue] (-3,-3.5) node{$\left(\widehat{p}^{(S)}_{i}\right)_{1\leq i\leq n_{\text{Test}}}$};
\draw[blue,->,>=latex] (-3.75,-2) -- (-3.75,-3) ;
\draw[purple,->,>=latex] (-5,-3.5) -- (-4.3,-3.5);
% S
\draw[ultra thick,black] (-7.575,2.25) rectangle (-1.25,-7.25);
\draw (-1.25,-7.25) node[above left]{\scriptsize$S\in\{a,b,\ldots,9\}$};

% Cadre de la couche
\draw[ultra thick,gray,dashed] (-7.75,2.5) rectangle (-1,-8);
\draw[gray] (-7.75,-8) node[above right]{\scriptsize $1^{\text{st}}$ layer};

%%% Couche 2
% Theta 2
\draw[purple] (-6,-9) node{$\widehat{\boldsymbol{\theta}}^{(2)}$};
\draw[purple,thick] (-6.5,-8.5) rectangle (-5.5,-9.5);
\draw[purple,->,>=latex] (-6,-7) -- (-6,-8.5) ;
% Y chapeau
\draw[purple!75!blue] (-3,-9) node{$\left(\widehat{P}_i\right)_{1\leq i\leq n_\text{Test}}$};
\draw[purple!75!blue,thick] (-4.5,-8.5) rectangle (-1.5,-9.5);
\draw[purple,->,>=latex] (-5.5,-9) -- (-4.5,-9) ;
\draw[purple!50!blue,->,>=latex] (-3.75,-4) -- (-3.75,-8.5) ;
% Cadre
\draw[ultra thick,gray,dashed] (-6.75,-8.25) rectangle (-1,-10);
\draw[gray] (-6.75,-10) node[above right]{\scriptsize $2^{\text{nd}}$ layer};

%%% Critere
\draw[purple!38!blue,thick] (3.5,-9) node{AUC};
\draw[purple!38!blue,thick] (1.5,-8.25) rectangle (5.5,-9.75);
\draw[blue,->,>=latex] (3.5,-5.5) -- (3.5,-8.25) ;
\draw[purple!75!blue,->,>=latex] (-1.5,-9) -- (1.5,-9) ;

% Big V-fold
\draw[ultra thick] (-8,-10.5) rectangle (7.25,3);
\draw (-8,-10.5) node[above right]{\small $5$-Fold};

    \end{tikzpicture}
% \hspace*{-2cm}
% \includegraphics[scale = 1]{images/schema_proc.pdf}
        \caption{Schematic representation of the procedure}
    \label{fig:schema:composition}
\end{figure}

A schematic representation is given in figure~\ref{fig:schema:composition}. The Raw data are composed of the coordinates of the handwriting trace recorded at $200 Hz$ of individual symbols written by the 545 children, where 479 ($88\%$) are typically developping children and 66 ($12\%$) have dysgraphia. The height of the gray box and the white box on the top represent the proportion of children with dysgraphia compared to children without dysgraphia in the database. Given a distance $\mathcal{L}_{\ell}$, through POMH the raw data are converted to feature table (Table~\ref{tab:data_secondlayermodel_bysymbol}). The big box at the bottom illustrates the 5-fold Cross Validation procedure. The individuals are divided into five folds by respecting the proportion of dysgraphia vs. no dysgraphia. At each time, $80\%$ of the data are used as training data and $20\%$ as test data. On the training data (workflow in red), in the first layer (left-top box inside the 5-fold box), the parameters of the first layer algorithm $\widehat{\boldsymbol{\theta}}^{(1)}_{S}$ and the output $\left(\widehat{p}^{(S)}_{i}\right)_{1\leq i\leq n_{\text{Test}}}$ are estimated. Then using the later as input, in the second layer (left-bottom box), the parameters of the second layer algorithm $\widehat{\boldsymbol{\theta}}^{(2)}_{S}$ are estimated. On the test data est (workflow in blue and purple), by applying the first layer algorithm with parameters $\widehat{\boldsymbol{\theta}}^{(1)}_{S}$ and the second layer algorithm with $\widehat{\boldsymbol{\theta}}^{(2)}_{S}$, the probability of dysgraphia at symbol level $\left(\widehat{p}^{(S)}_{i}\right)_{1\leq i\leq n_{\text{Test}}}$ and at individual level $\left(\widehat{P}_i\right)_{1\leq i\leq n_\text{Test}}$ are estimated. In the end, by applying cut-off value on $\left(\widehat{P}_i\right)_{1\leq i\leq n_\text{Test}}$, we obtain $\left(\widehat{Y}_i\right)_{1\leq i\leq n_\text{Test}}$. By comparing it with $\left(Y_i\right)_{1\leq i\leq n_\text{Test}}$, the AUC value on one Train-Test set is calculate. The final AUC is the average of the 5-fold.

% \begin{figure}[!ht]
%     \centering
%     \includegraphics[width=0.8\linewidth]{Draft-graph.png}
%     \caption{\VB{A faire en enlevant du coup les "S" dans la première couche}}
%     \label{fig:enter-label}
% \end{figure}
\begin{table}[!ht]
    \centering
    \caption{Summary of method combinations used between the first (in row) and second layers (in column).}
    \label{tab:summarise:layers}
    \begin{tabular}{|c|c||* {3}{c|}}
    \cline{3-5}
         \multicolumn{2}{c|}{}&\multicolumn{3}{c|}{$2^{\text{nd}}$  layer}\\
    \cline{3-5}
         \multicolumn{2}{c|}{}&Counting&GLM&RF\\
    \hline
    \multirow{3}{*}{\rotatebox{90}{$1^{\text{st}}$ layer}}&GLM& \checkmark& \checkmark&\\
    \cline{2-5}
    &RF& \checkmark&& \checkmark\\
    \cline{2-5}
    &SVM& \checkmark&&\\
    \hline
    \end{tabular}
\end{table}

\section{Comparison of the results from the five classifiers}\label{sec:res}
The procedure of the two-layer classifier and five classifiers of different methods combinations are presented in the previous section. In this section, we detail the parametrization of each classfier and compare the results.
\subsection{GLM~$\xrightarrow{}$~Counting}

\paragraph{Model specification}
In order to simplify the modeling and reporting of results, variables \texttt{section}, \texttt{total-time} and $w_x$ are dichotomized into discrete variables. As a consequence, \texttt{gr$\_$sec} is introduced with levels \texttt{1\textsuperscript{st} cycle (1st- to 4th-grade) and 2\textsuperscript{nd} cycle (5th- to 9th-grade)} which are different grades in the french educational system, while \texttt{gr$\_$speed} has two levels \texttt{fast} or \texttt{slow}, indicating whether the time is respectively longer or shorter than the average time. The operator size \texttt{gr$\_$wx} has levels \texttt{small} ($3 \leq w_x \leq \text{median}(w_x)$) and \texttt{large} ($\text{median}(w_x) \leq w_x \leq 39$). Therefore, among the predictors, we keep one continuous variable of interest (\texttt{dist}) and three categorical variables with two levels (\texttt{gr$\_$sec}, \texttt{gr$\_$wx} and \texttt{gr$\_$speed}). %While specifying the GLM model, the effect of individual covariate and factors and the interaction effect between the covariate and factors are considered. 

%\vspace{0.5cm}
%\paragraph{Interaction effect}
%An interaction effect is a way of specifying that the relationship between the outcome $Y$ and a predictor $X_1$ can be dependent on another predictor $X_2$.

While the main purpose of the modeling is to study the effect of \texttt{dist} on the detection of dysgraphic handwriting, this effect should be dependent on the value of other factors. For instance, in the group with smaller and greater $w_x$, \texttt{dist} does not have the same effect on the output; the probability that a symbol is written by someone with dysgraphia increases faster with \texttt{dist} in older children than in younger children, since the latter are still in the learning phase. We therefore included all interaction terms between \texttt{dist}, \texttt{gr$\_$sec}, \texttt{gr$\_$speed} and \texttt{gr$\_$wx}, thus including the three two-way and the three three-way interactions.

\vspace{0.5cm}

%L'intercept, je trouve normal qu'on le voit puisque, naturellement, les enfants ne sont pas dysgraphiques (Vincent B.). 

%Par exemple pour la lettre z, l'age n'est pas significative pas l'interaction entre l'age et la vitesse de l'ecriture (le temps total) est significative. Ainsi dist et la taille de l'operateur (wx).

\paragraph{Results}

\begin{figure}
\vspace{-1cm}
    \centering
    \includegraphics[width = \textwidth]{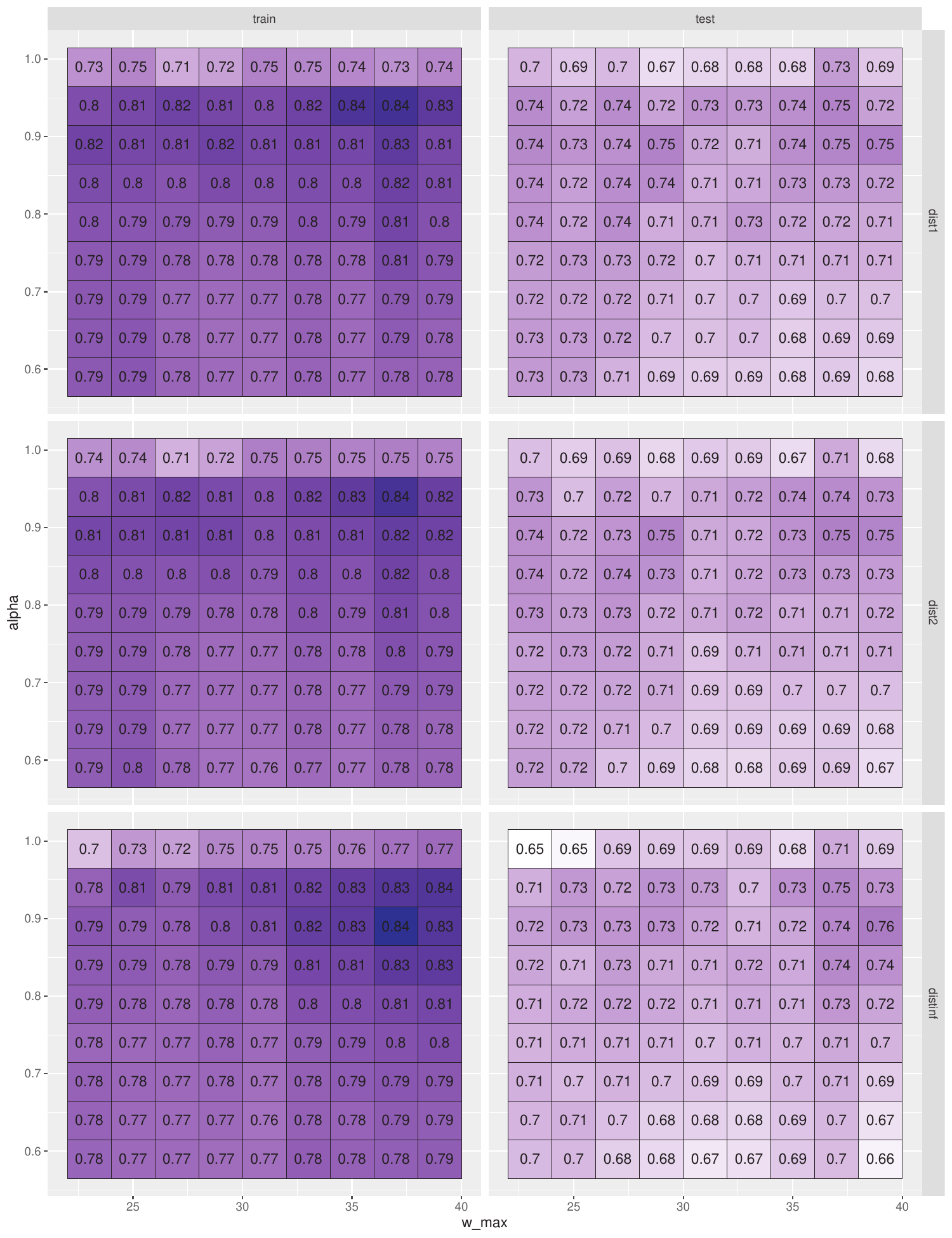}
    \caption{Classifier GLM~$\xrightarrow{}$~Counting: AUC value according to \texttt{max$\_$w} (table columns) and \texttt{$\alpha$} (table rows) for the train and test sets (facet columns), with different distance types (facet rows).}
    \label{fig:auc_glm_pi}
\end{figure}

The AUC (air under curve) with different values of threshold $\alpha$ (see Eq. \ref{eq:count_alpha}) and $w_\text{max}$ is shown in Fig.~\ref{fig:auc_glm_pi}, for $\mathcal{L}_1$, $\mathcal{L}_2$, and $\mathcal{L}_\infty$ respectively. The AUC value is the average value by 5-folds cross-validation. It is noticed that the largest AUC values by column are located in the middle instead of on the edge, which means that 39 is large enough to make sure that the optimal $w_\text{max}$ is located between 23 and 39.

\begin{figure}
    \centering
    \includegraphics[width = \textwidth]{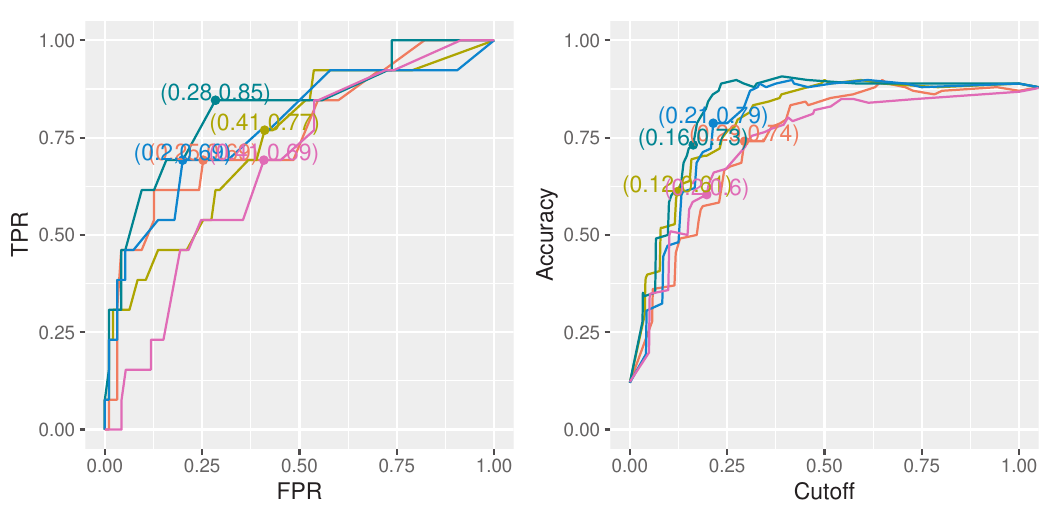}
    \caption{GLM~$\xrightarrow{}$~Counting: ROC curves giving the maximum AUC value and the corresponding accuracy curve. Colors correspond to the five iterations.}
    \label{fig:fpr_tpr_glm_pi_te}
\end{figure}

Table~\ref{tab:res_glm_pi} summarises the maximum AUC values on test datasets for each type of distance and the optimal parameters. Comparing three types of distances, $\mathcal{L}_\infty$ gives the best result, but the difference is very small. If we zoom in at that point ($\mathcal{L}_\infty$, $\alpha$ = 0.89, $w_\text{max} = 39$), the FPR, TPR and the accuracy are presented in Fig.~\ref{fig:fpr_tpr_glm_pi_te}. Colors correspond to iterations. The points tagged on ROC curves (FPR vs TPR) corresponds to the nearest point on the curve to point (0,1). The same points are tagged on the accuracy plot as well, where the cut-off value means that if the probability of an individual is greater than the value then it is classified as having dysgraphia and otherwise not having dysgraphia. By taking the average of the 5 folds, it gives FPR = 0.31, TPR = 0.73, cut-off = 0.20, ACC = 0.69, where cut-off=0.20 means that if the individual has more than 7 ($0.20\times36$) symbols classified as dysgraphic, then it is considered as having dysgraphia.

\begin{table}[ht]
\caption{GLM~$\xrightarrow{}$~Counting: Optimal parameters for three types of distance} 
\centering
\begin{tabular}{rlrrrr}
  \hline
 & ty\_dist & w\_max & alpha & auc\_train & auc\_test \\ 
  \hline
1 & dist1 & 39 & 0.890 & 0.810 & 0.749 \\ 
  2 & dist2 & 37 & 0.890 & 0.824 & 0.751 \\ 
  3 & distinf & 39 & 0.890 & 0.834 & \cellcolor{red!20} 0.755 \\ 
   \hline
\end{tabular}

\label{tab:res_glm_pi}
\end{table}

\subsection{GLM~$\xrightarrow{}$~GLM}
\begin{figure}
    \centering
    \includegraphics[width = \textwidth]{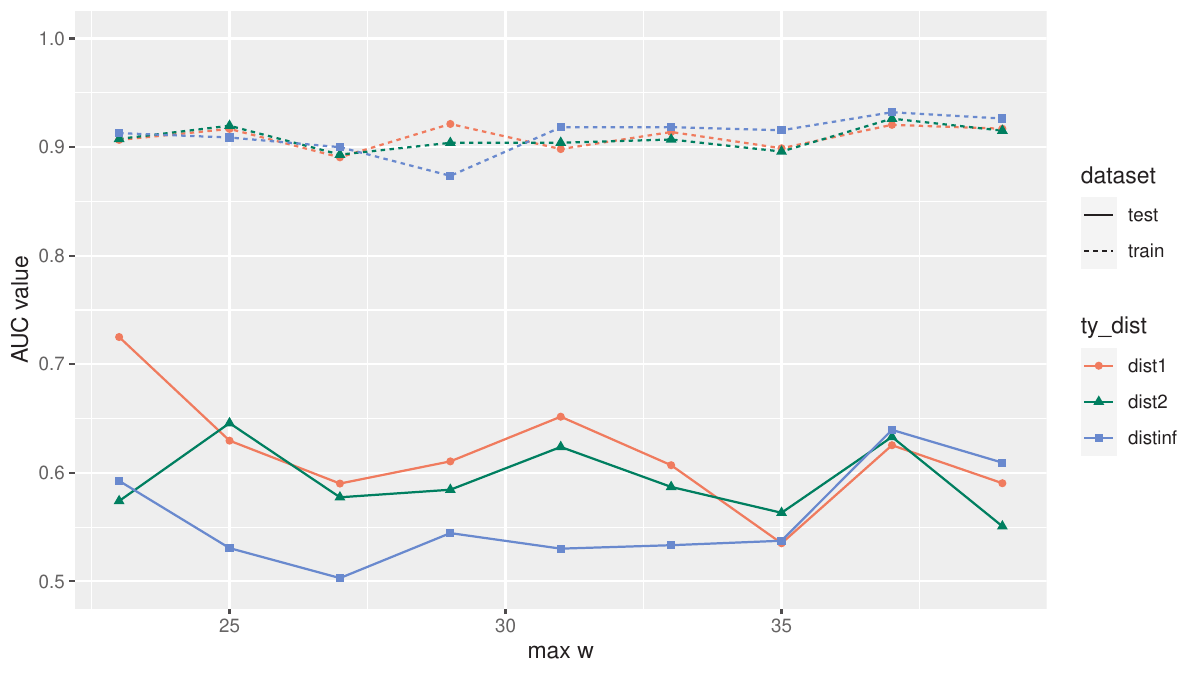}
    \caption{GLM~$\xrightarrow{}$~GLM: The AUC value as function of $w_\text{max}$}
    \label{fig:auc_glm_glm}
\end{figure}
The fitted values from GLM models in the first layer, $\widehat{p}_{i}^S, S = {a, b, \cdots, 9}, i = 1, 2, \cdots, n$, are used as explanatory variables of the second GLM model. If individual $i$ didn't write symbol $S$, then $\widehat{p}_{i}^S$ is missing. We choose to delete the record if an individual didn't write all the 36 symbols, which results in the loss of $24.2\%$ TD and $36.4\%$ dysgraphic observations (Table~\ref{tab:widgets1}). In the second layer, no interaction effect is considered. The variables are selected backward by the AIC criterion. The fitted value of the model in the second layer $\widehat{P}_i, i = 1, \cdots, n$ indicates the probability that an individual displays dysgraphia.

The AUC value is calculated by drawing the ROC curve when varying the cut-off value on $\widehat{P}_i, i = 1, \cdots, n$ from 0 to 1. It is noticed that the difference in the results between the training data and test data are not negligible (Fig.~\ref{fig:auc_glm_glm}). The AUC value for the training data is nearly $>0.9$ for all $w_\text{max}$. However the AUC values on the test set are much lower, $0.725, 0.646$ and $0.640$ using $\mathcal{L}_1$, $\mathcal{L}_2$ and $\mathcal{L}_\text{inf}$ distances respectively (Table~\ref{tab:res_glm_glm}). It shows that there is an overfitting problem. The details of TPR, FPR and Accuracy under the optimal model with distance $\mathcal{L}_1$, $w_\text{max}=23$ can be found in Fig.~\ref{fig:fpr_tpr_glm_glm_te} Supplementary Materials.

\begin{table}[ht]
\centering
\caption{GLM~$\xrightarrow{}$~GLM: Optimal parameters for three types of distance} 
\begin{tabular}{rlrrr}
  \hline
 & ty\_dist & w\_max & auc\_train & auc\_test \\ 
  \hline
1 & dist1 & 23 & 0.907 & \cellcolor{red!20} 0.725 \\ 
  2 & dist2 & 25 & 0.920 & 0.646 \\ 
  3 & distinf & 37 & 0.932 & 0.640 \\ 
   \hline
\end{tabular}
\label{tab:res_glm_glm}
\end{table}

\subsection{RF~$\xrightarrow{}$~Counting}

The package \texttt{randomForest} is used to build the Random Forest classifier. In the first layer RF, the predictors and target variable are similar as in the first layer GLM. In total 36 RF models are built, each for one symbol. As explained in Section~\ref{subsec:estimate_w}, $\widehat{w}_x$ and $\widehat{w}_y$ are the estimates of the "closing" operator size on two orthogonal directions. As the two estimates are related to the intrinsic writing regularity of a person, they are highly correlated. Because each tree is grown on a sampled number of predictors, Decision Trees are not correlated among them. The algorithm is therefore robust towards correlated variables, in other words collinearity is not a problem in RF algorithm. We include both $w_x$ and $w_y$ to gain a higher prediction ability of the model. The predictors included in each model are \texttt{dist},\texttt{section}, \texttt{total-time}, $w_x$ and $w_y$. We set \texttt{ntree = 500} (number of trees to grow), \texttt{mtry = 2} (number of variables randomly sampled as candidates for split) in function \texttt{randomForest()}. To clarify, as bootstrap is used in RF when multiple trees are grown, no CV are needed inside the first layer.

The probability $\widehat{p}_i^S$ is calculated by the proportion of votes where $Y_{i}^S = 1$. Then the Counting method described above is used to obtain AUC value both on train and test datasets.

%Which criteria is used in function \texttt{randomForest()} to grow one tree, is it Gini or entropy? GINI (The choice of cut.off value or the thresholding method on second layer don't influence the construction of trees.)

%Probability = Vote ((classification only) a matrix with one row for each input data point and one column for each class, giving the fraction or number of (OOB) ‘votes’ from the random forest.)
%(cut.off is a parameter for individual decision tree, its values in most of the cases don't make sense)

\begin{figure}
\vspace{-1cm}
    \centering
    \includegraphics[width = \textwidth]{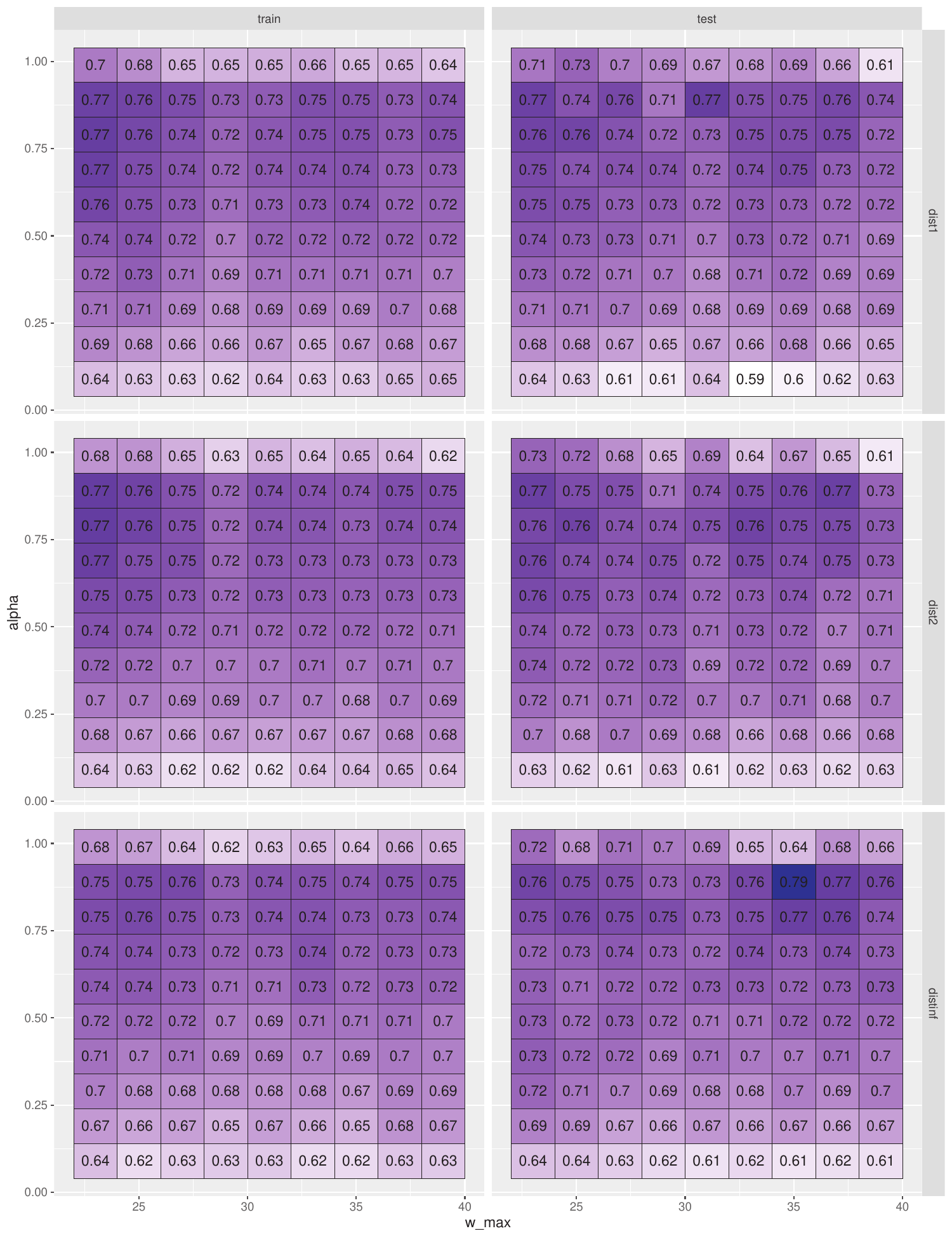}
    \caption{Classifier RF~$\xrightarrow{}$~Counting: AUC value according to \texttt{max$\_$w} (table columns) and \texttt{$\alpha$} (table rows) for the train and test sets (facet columns), with different distance types (facet rows).}
    \label{fig:auc_rf_pi}
\end{figure}

It can be noticed in Fig.~\ref{fig:auc_rf_pi} that, in general, for all types of distance and for both train and test datasets, $\alpha = 0.89$ (the second line in each tile) gives the highest auc value. A summary of the best parameters for three types of distance is given in Table~\ref{tab:res_rf_pi}. According to the AUC value on test datasets, optimal combination of parameters, $\mathcal{L}_{\infty}, w_\text{max}, \alpha = 0.89$ gives maximum AUC value of $0.792$. It can be noticed that AUC values in test datasets are slightly higher than in train datasets with the same parameters. This phenomenon is supposed to be caused by random factors. 

\begin{table}[ht]
\centering
\caption{RF~$\xrightarrow{}$~Counting: Optimal parameters for three types of distance} 
\begin{tabular}{rlrrrr}
  \hline
 & ty\_dist & w\_max & alpha & auc\_train & auc\_test \\ 
  \hline
1 & dist1 & 31 & 0.890 & 0.734 & 0.771 \\ 
  2 & dist2 & 37 & 0.890 & 0.750 & 0.768 \\ 
  3 & distinf & 35 & 0.890 & 0.739 & \cellcolor{red!20}0.792 \\ 
   \hline
\end{tabular}
\label{tab:res_rf_pi}
\end{table}

\subsection{RF~$\xrightarrow{}$~RF}

Similar to the second layer GLM, the second layer RF uses the output of the first layer RF, $\widehat{p}_{i}^S, S = {a, b, \cdots, 9}, i = 1, 2, \cdots, n$, as predictors. The missing values are omitted in the same way. The output of the second layer $\widehat{q}_i, i = 1, \cdots, n$ indicates the probability that an individual is detected as having dysgraphia. When implementing the second layer RF by function \texttt{randomForest}, the arguments are set to \texttt{ntree} = 500, \texttt{mtry} = 6.

\begin{figure}[!ht]
    \centering
    \includegraphics[width = \textwidth]{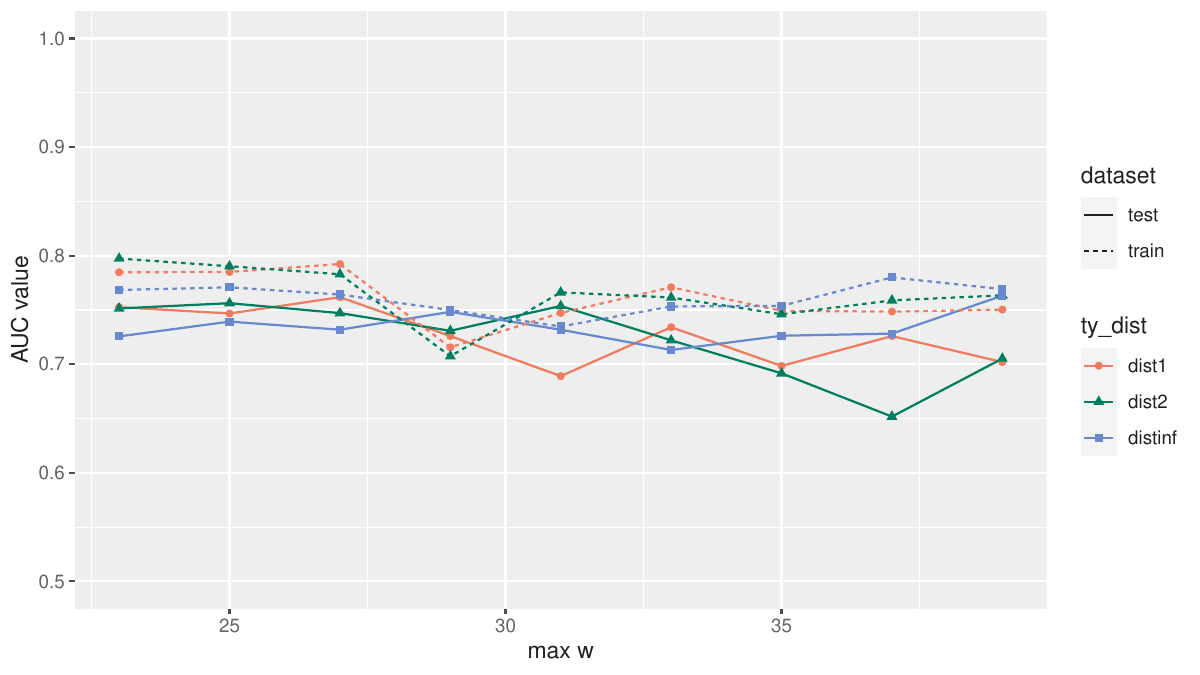}
    \caption{Classifier RF~$\xrightarrow{}$~RF: The AUC values as function of $w_\text{max}$ with distance $\mathcal{L}_1, \mathcal{L}_2$ and $\mathcal{L}_\infty$}
    \label{fig:auc_rf_rf}
\end{figure}

The AUC values in test dataset are as good as in train datasets (Fig.~\ref{fig:auc_rf_rf}). The best AUC value is given by $\mathcal{L}_{\infty}, w_\text{max}=39$ (Table~\ref{tab:res_rf_rf}).
\begin{table}[ht]
\centering
\caption{
: Optimal parameters for three types of distance} 
\begin{tabular}{rlrrr}
  \hline
 & ty\_dist & w\_max & auc\_train & auc\_test \\ 
  \hline
1 & dist1 & 27 & 0.792 & 0.762 \\ 
  2 & dist2 & 25 & 0.790 & 0.756 \\ 
  3 & distinf & 39 & 0.769 &\cellcolor{red!20} 0.763 \\ 
   \hline
\end{tabular}
\label{tab:res_rf_rf}
\end{table}

\subsection{SVM~$\xrightarrow{}$~Counting}\label{subsec:svm+counting}
The package \texttt{e1071} is used to build SVM models \cite{dimitriadou2006e1071,meyer2015support}. As the two classes are not balanced, option \texttt{class.weights = "inverse"} is set in function \texttt{svm()} to choose the weights inversely proportional to the class distribution. The RBF kernel is used because of its versatility and because there are fewer parameters, $\gamma$ and $C$, to tune compared to other kernels (e.g. polynomial). The optimal parameters, $\gamma$ and $C$, are learned by an inner 5-fold Cross Validation, using classification error as metric, by grid search in the range of $\gamma \in [2^{-10}, 2^{-2}]$ and $C\in [2^5,2^{10}]$. It has been checked that optimal parameters lay most of the time inside the range and not on the border of the range. As it was approved by many previous studies  \cite{dimitriadou2006e1071,meyer2015support}) that scaling of the predictors can in most of the cases drastically improve the result. It is therefore applied in our algorithm settings.

% {\color{red} To check. Another related question, how probability is calculated in \texttt{svm()} and why it is so unstable} The value of $wx_i+b$ is not interpretable, but only the sign of it determines which side of the hyperplane the point is attributed. (\VB{Reference? why it is not interpretable??})
% $f(x_i) = \text{sign}(wx_i + b)$. Therefore, the output of SVM is only binary and the model does not give the probability itself, unless when a cross-validation framework is used. 

Although in the function \texttt{svm}, there is the option to choose probability as output of the algorithm, no information is known to the authors about how the probability is calculated based on the SVM algorithm. One technique \cite{li2012power} proposes to use Logistic Regression to map SVM scores into probability. The option of probability is tested in the frame of SVM $\xrightarrow{}$ SVM (Appx. \ref{app:svm2}). However it has shown unstable results. In order to avoid the interpretation problem about probability, we stick into the binary classification results. In the second layer of the modeling, the detection of dysgraphia is done by calculating the proportion of positive symbols. The AUC value is calculated by varying the proportion from 0 to 1.

\begin{figure}
    \centering
    \includegraphics[width = \textwidth]{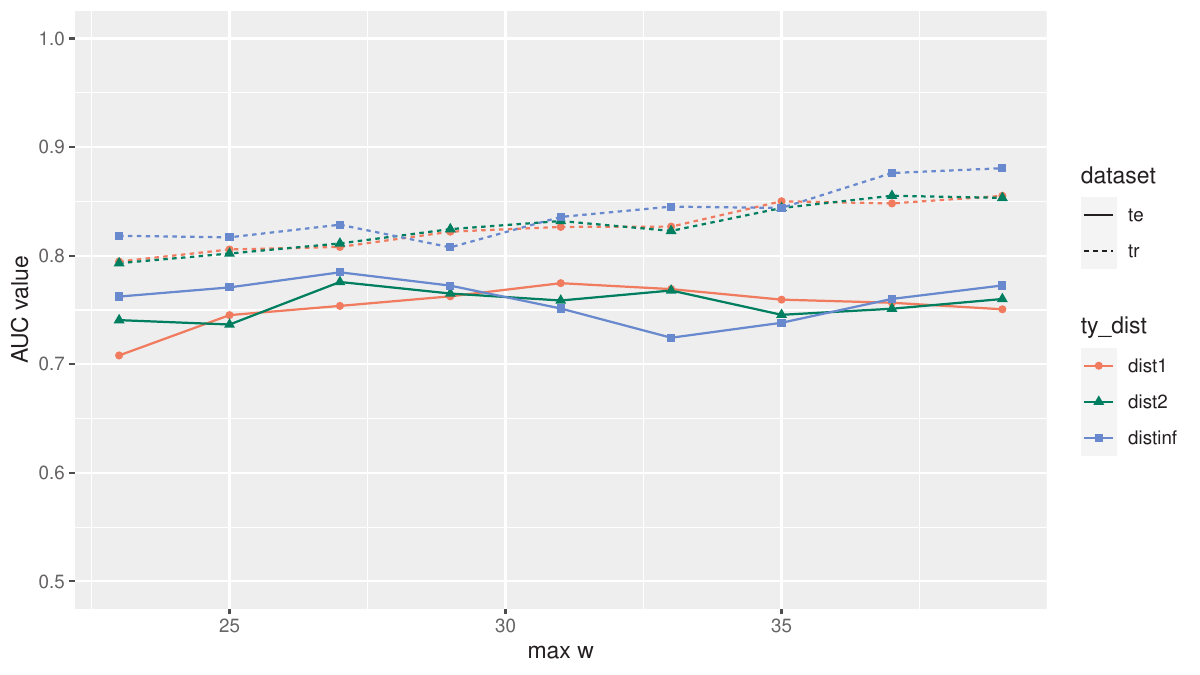}
    \caption{Classifier SVM $\xrightarrow{}$ Counting: The AUC values as function of $w_\text{max}$ with distance $\mathcal{L}_1, \mathcal{L}_2$ and $\mathcal{L}_\infty$}
    \label{fig:svm_count}
\end{figure}

Promising prediction results are shown in Fig.~\ref{fig:svm_count}. It can be noticed that the prediction results on test sets are quite stable across different $w_\text{max}$ and different distance types, with maximum value of $0.785$ at  $w_\text{max}=27$ and $\mathcal{L}_\infty$ distance (Table~\ref{tab:res_svm_nb}).

\begin{table}[ht]
\centering
\begin{tabular}{rlrrr}
  \hline
 & ty\_dist & w\_max & auc\_train & auc\_test \\ 
  \hline
1 & dist1 & 31 & 0.827 & 0.775 \\ 
  2 & dist2 & 27 & 0.811 & 0.776 \\ 
  3 & distinf & 27 & 0.829 & \cellcolor{red!20}0.785 \\ 
   \hline
\end{tabular}
\caption{SVM $\xrightarrow{}$ Counting: Optimal parameters for three types of distance} 
\label{tab:res_svm_nb}
\end{table}

\paragraph{Synthesis of the results}
To summarize, the classifier RF~$\xrightarrow{}$~Counting gives the highest AUC value on the test dataset $0.792$ (Table~\ref{tab:res_rf_pi}), followed by SVM~$\xrightarrow{}$~Counting (0.785 Table~\ref{tab:res_svm_nb}), then RF~$\xrightarrow{}$~RF (0.763 Table~\ref{tab:res_rf_rf}) and GLM~$\xrightarrow{}$~Counting (0.755 Table~\ref{tab:res_glm_pi})). Overfitting problem is noticed in GLM~$\xrightarrow{}$~GLM, where the AUC on the test datasets are much lower than on the training datasets (Table~\ref{tab:res_glm_glm}). It has also been noticed that, for classifier RF~$\xrightarrow{}$~Counting, the AUC values on the test datasets are higher than on the training datasets. This unusual fact may be due to random factors.

As explained in Section~\ref{sec:nbzero_vs_age}~and~\ref{subsec:estimate_w}, the parameter $w_\text{max}$ has a direct impact on the estimates $\hat{w}$. Empirically, we created feature datasets with $w_\text{max}$ ranging from 23 to 39. Through the two-layer procedure and CV, the optimal $w_\text{max}$ is chosen by each classifier. No rule has been found on how parameter $w_\text{max}$ influences the AUC value. The fact that the relationship between $w_\text{max}$ and AUC is not obvious can be explained by the complexity of the combination of different models, Mathematical Morphology, POMH and two-layer classification. In most of the cases, the optimal $w_\text{max}$ falls in the middle of the interval $[23, 39]$, not on the edge, which means that the interval is well chosen to include the optimal value. However there are some exceptions, e.g. in GLM~$\xrightarrow{}$~GLM (Table~\ref{tab:res_glm_glm}) where $w_\text{max}=23$.

We were interested to know if different types of distance, when calculating the distance between the original and reconstructed handwriting traces (Section~\ref{subsec:ty_dist}), influence the prediction ability of the classifiers. Through the two-layer classifier and cross validation (CV), it shows that there is no one distance type better than the others in terms of the prediction ability of the classifier, except in GLM~$\xrightarrow{}$~GLM (Table~\ref{tab:res_glm_glm}) where $\mathcal{L}_1$ gives the largest AUC value.

In the classifier GLM $\xrightarrow{}$ Counting and RF $\xrightarrow{}$ Counting, after the first layer, there is the threshold $\alpha$ (Eq.~\ref{eq:count_alpha}) to determine whether a symbol is written by a participant with dysgraphia. For both classifiers, by the CV procedure, the optimal $\alpha$ is $0.89$ (Tables~\ref{tab:res_glm_pi}~and~\ref{tab:res_rf_pi}) which is way higher than $0.5$. If the default value 0.5 was used, the prediction ability of the classifier will be compromised.

\section{Discussion and Perspectives}

\subsection{Discussion}
The two-layer classifiers give promising classification results. The prediction ability is higher compared to the original one-layer classification model. The results are intuitively explainable. 

It is worth noticing that the factor of age is included in the modeling, both in the estimation of $w$, the operator size and in the two-layer classifier. There is one issue that the two classes in the sample are imbalanced. By thresholding the probability $\widehat{p}^{(S)}_{i}$ by its $\alpha$ order quantile instead of using $0.5$ for GLM and RF, by setting \texttt{class.weights = "inverse"} in SVM, and by maximizing AUC value instead of accuracy, the classifers alleviate the problem of the imbalance of classes.

For classifiers GLM~$\xrightarrow{}$~Counting,  RF~$\xrightarrow{}$~Counting and SVM~$\xrightarrow{}$~Counting, on the training dataset, if we choose the point on ROC curve the closest to point $(0,1)$ as the optimal, then the cut-off value, on the proportion of positive symbols from the first layer model, is $0.20$, $0.311$, $0.272$, respectively. It means that if an individual wrote all 36 symbols,, when more that 7, 11 or 9 (for the three models respectively) are estimated as positive numbers, the individual is classified as having dysgraphia. 

An advantage of using Counting in the second layer is that all individuals can be included into modeling even those who didn't write all symbols. On the contrary, individuals with missing symbols are excluded in GLM and RF on the second layer.

% \jcq{Premier jet de ce bout de discussion sur les limites du filtrage et les extensions ou généralisations possibles. A compléter ou améliorer, sachant que je ne suis pas certain que ce soit les meilleurs points pour terminer l'article, même si "acceptable".}

% Analyse of the integration of age in the procedure
% The model deal with the inbalance of the two classes
% See how detection False negative and False positive are related to other neurological conditions
\subsection{Limitations/perspectives} 
\paragraph{Denoising Approaches} Although sufficient to demonstrate the relevance of our approach, the method based on morphological operators that was used to preprocess raw data and select zero velocity points remains sensitive to the window size parameter. Since POMH models the velocity of handwriting traces, we focused at this level to perform the filtering steps and tested the stability of the results. However, filtering coordinates before derivation should likely improve the signal-to-noise ratio. Indeed, the complex dynamics of pen and paper (on tablet) interactions may lead to strong non-linearity and spurious detection of zero points, which may be harder to filter out on velocity and acceleration compared to spatial coordinates; friction forces for instance, especially static friction, must be overcome when starting any pen movement and moving over a rough writing surface.

Adopting entirely different preprocessing methods would alleviate the need to tuning the morphological parameters, while probably introducing other parameters. Fourier transform has often been used in the literature (\cite{mueller2009human}, \cite{velay2013signal}), as filtering out high frequency components eliminates small and rapid movements resulting from jerk in motor control or from friction related events. 

% \jcq{Demander à Caroline la référence la plus solide, vu que c'est elle qui avait rappelé l'usage répandu de cette méthode ; je sais que j'étais parti moi aussi là-dessus dès 2007, mais je suis rouillé des références depuis le temps...}

In the presented modeling approach, we opted for separated filtering and POMH parameter estimation steps. However and to go even further to eliminate dependency to filtering parameters, dynamic time-warping or dynamic programming on velocity profiles could be used to estimate POMH parameters while maximising the fit to raw noisy data. Algorithmic complexity would of course increase, and mathematical developments and optimization would be required, but such methods should remain tractable on individual letters or short sequences.

In any case, especially using the generic and low complexity filtering method introduced in the article, our multi-step approach could be applied to many kinds of functional and spatiotemporal sampled data. Also, abstracting from POMH, the benchmarking of combined algorithms could therefore be generalized or applied to other multi-level classification problems, beyond the assisted diagnosis of dysgraphia.

\paragraph{Choice of cut-off value on AUC curve}
When we choose the point on ROC curve the closest to point $(0,1)$ as the optimal, accordingly, the FPR and TPR can be obtained. In clinical practice, there can be different ways of choosing the cut-off value, for example, by setting a fixed FPR or a desired TPR. The choice should be validated on the test dataset. This problem is out of the scope of this article.

\begin{appendix}
\section{Appendix}
\subsection{Distribution of number of zeros with different $w_\text{max}$}
\begin{figure}[!t]
    \centering
    \includegraphics{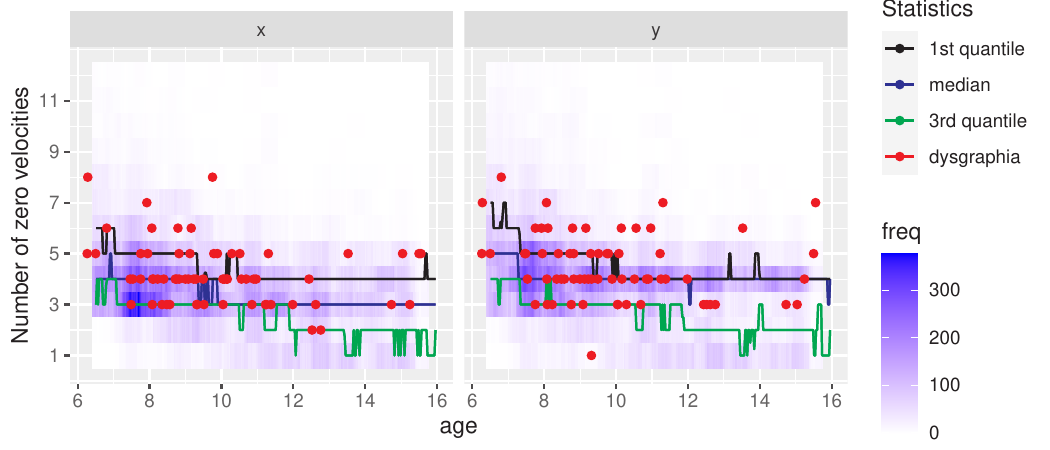}
    \caption{Density and statistics of the number of zeros on $x$ and $y$-axis against age, for letter "a", $mw=39$}
    \label{fig:enter-label}
\end{figure}
\subsection{Example of a first layer GLM model}

\begin{lstlisting}[style=Rstyle]
mod_full <- glm(data = df_mod, formula = class ~ dist_norm + gr_sec + gr_wx + gr_slow + dist_norm : gr_wx + dist_norm : gr_sec + dist_norm : gr_slow + dist_norm : gr_slow : gr_wx + dist_norm : gr_slow : gr_sec + dist_norm : gr_sec : gr_wx, family = "binomial")      
\end{lstlisting}

\vspace{0.5cm}
\noindent \textbf{Variables Selection}

\noindent \textbf{Stepwise:}
\begin{lstlisting}[style=Rstyle]
mglm <- step(mod_full, trace = 0, direction = "both")
\end{lstlisting}
\noindent \textbf{AIC:}
\begin{lstlisting}[style=Rstyle]
mglmAIC <- MASS::stepAIC(mod_full, trace = 0, k = 2)
\end{lstlisting}
\noindent \textbf{BIC:}
\begin{lstlisting}[style=Rstyle]
modBIC <- MASS::stepAIC(mod, trace = 0, k = log(nrow(df_mod)))
\end{lstlisting}

\begin{table}[ht]
\centering
\caption{coefficients of model for letter a} 
\label{tab:coef_a_before_trans}
\begin{tabular}{rlrrrr}
  \hline
 & var & coef & std\_error & z\_value & p\_value \\ 
  \hline
3 & (Intercept) & -2.05 & 0.30 & -6.91 & 0.00 \\ 
  4 & dist\_norm & 0.52 & 0.45 & 1.16 & 0.25 \\ 
  5 & gr\_seccycle2 & -0.99 & 0.43 & -2.29 & 0.02 \\ 
  6 & gr\_wxTRUE & 0.85 & 0.40 & 2.12 & 0.03 \\ 
  7 & gr\_slowTRUE & -0.51 & 0.44 & -1.17 & 0.24 \\ 
  8 & dist\_norm:gr\_wxTRUE & 1.34 & 0.73 & 1.83 & 0.07 \\ 
  9 & dist\_norm:gr\_seccycle2 & -1.08 & 0.76 & -1.42 & 0.15 \\ 
  10 & dist\_norm:gr\_slowTRUE & -1.28 & 0.78 & -1.64 & 0.10 \\ 
  11 & dist\_norm:gr\_seccycle2:gr\_slowTRUE & 3.54 & 1.40 & 2.52 & 0.01 \\ 
   \hline
\end{tabular}
\end{table}

\begin{table}[ht]
\tiny
\centering
\caption{coefficients of model for letter "a"} 
\label{tab:coef_a}
\begin{tabular}{rlllllllll}
  \hline
 & ord3\_var & ord3\_sig & ord3\_pos & ord2\_var & ord2\_sig & ord2\_pos & ord1\_var & ord1\_sig & ord1\_pos \\ 
  \hline
1 & dist\_norm:gr\_seccycle2:gr\_slowTRUE & 1 & 1 & dist\_norm:gr\_seccycle2 & 0 & 0 & dist\_norm & 0 & 1 \\ 
\rowcolor{red!20} % Change the background color of the second row
  2 & dist\_norm:gr\_seccycle2:gr\_slowTRUE & 1 & 1 & dist\_norm:gr\_seccycle2 & 0 & 0 & gr\_seccycle2 & 1 & 0 \\ 
  3 & dist\_norm:gr\_seccycle2:gr\_slowTRUE & 1 & 1 & dist\_norm:gr\_slowTRUE & 0 & 0 & dist\_norm & 0 & 1 \\ 
  4 & dist\_norm:gr\_seccycle2:gr\_slowTRUE & 1 & 1 & dist\_norm:gr\_slowTRUE & 0 & 0 & gr\_slowTRUE & 0 & 0 \\ 
  5 & dist\_norm:gr\_seccycle2:gr\_wxTRUE & 0 & 0 & dist\_norm:gr\_seccycle2 & 0 & 0 & dist\_norm & 0 & 1 \\ 
  6 & dist\_norm:gr\_seccycle2:gr\_wxTRUE & 0 & 0 & dist\_norm:gr\_seccycle2 & 0 & 0 & gr\_seccycle2 & 1 & 0 \\ 
  7 & dist\_norm:gr\_seccycle2:gr\_wxTRUE & 0 & 0 & dist\_norm:gr\_wxTRUE & 1 & 1 & dist\_norm & 0 & 1 \\ 
  8 & dist\_norm:gr\_seccycle2:gr\_wxTRUE & 0 & 0 & \cellcolor{blue!20}dist\_norm:gr\_wxTRUE & \cellcolor{blue!20}1 & \cellcolor{blue!20}1 & \cellcolor{blue!20}gr\_wxTRUE & \cellcolor{blue!20}1 & \cellcolor{blue!20}1 \\ 
  9 & dist\_norm:gr\_wxTRUE:gr\_slowTRUE & 0 & 0 & dist\_norm:gr\_wxTRUE & 1 & 1 & dist\_norm & 0 & 1 \\ 
  10 & dist\_norm:gr\_wxTRUE:gr\_slowTRUE & 0 & 0 & dist\_norm:gr\_wxTRUE & 1 & 1 & gr\_wxTRUE & 1 & 1 \\ 
  11 & dist\_norm:gr\_wxTRUE:gr\_slowTRUE & 0 & 0 & dist\_norm:gr\_slowTRUE & 0 & 0 & dist\_norm & 0 & 1 \\ 
  12 & dist\_norm:gr\_wxTRUE:gr\_slowTRUE & 0 & 0 & dist\_norm:gr\_slowTRUE & 0 & 0 & gr\_slowTRUE & 0 & 0 \\ 
   \hline
\end{tabular}
\end{table}

%\todo[inline, color=red!40]{The way to discretize variables are questioned and worth further discussion.}%

%\todo[inline, color=red!40]{groupe d'age en fonction de cycle d'éducation francais, cycle 1: CP CE1 CE2, cycle 2 CM1 CM2, 6ème cycle 3 5ème - 3ème}%

%\todo[inline, color=red!40]{Un tableau de résumé: variable vs discriminative ou non pour des 36 symboles}%
\paragraph{Statistics on the coefficients for the 36 models}

The proportion of models where the third order interaction \texttt{dist$\_$norm:gr$\_$seccycle2:gr$\_$slowTRUE} (\texttt{dist$\_$norm:gr$\_$seccycle2:gr$\_$wxTRUE} and \texttt{dist$\_$norm:gr$\_$wxTRUE:gr$\_$slowTRUE} respectively) is significant is 0.2376543 (0.1450617 and 0.2407407 respectively). And in case the third order interaction \texttt{dist$\_$norm:gr$\_$seccycle2:gr$\_$slowTRUE} (\texttt{dist$\_$norm:gr$\_$seccycle2:gr$\_$wxTRUE} and \texttt{dist$\_$norm:gr$\_$wxTRUE:gr$\_$slowTRUE} respectively) is significant, the proportion of positive coefficient is $100\%$ ($0\%$ and $0\%$).

The proportion of the model where the intercept is negative is $100\%$.

When higher order interactions are not significant, then the coefficient of \texttt{dist} is always positive, else wise it may change to negative. 

%\textbf{Model prediction}
\subsection{GLM + GLM}
\begin{figure}
    \centering
    \includegraphics[width = \textwidth]{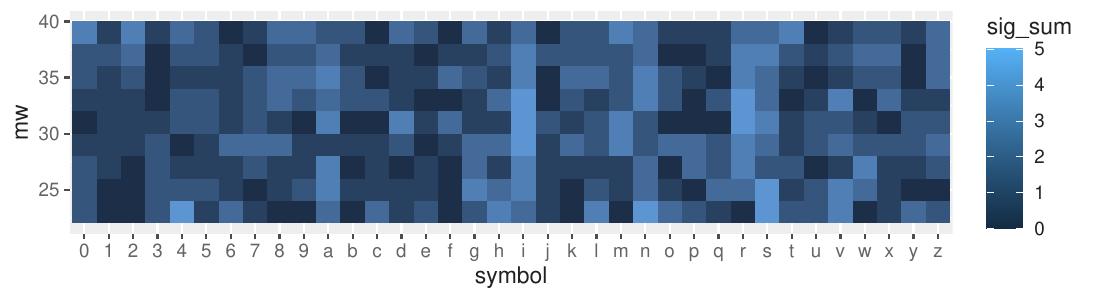}
    \caption{Count how many times among the 5 folds are the symbols significant by AIC criterion ($\mathcal{L}_1$ distance).}
    \label{fig:sig_num_glm_glm}
\end{figure}
It can be observed in Fig. \ref{fig:sig_num_glm_glm} that symbols like a, i, n, r and s are more often significant or discriminative to detect dysgraphia. 

\subsection{Cut-off value and the False Postive Rate (FPR) and True Positive Rate (TPR)}\label{app:fpr_tpr}
\begin{figure}
    \centering
    \includegraphics[width = \textwidth]{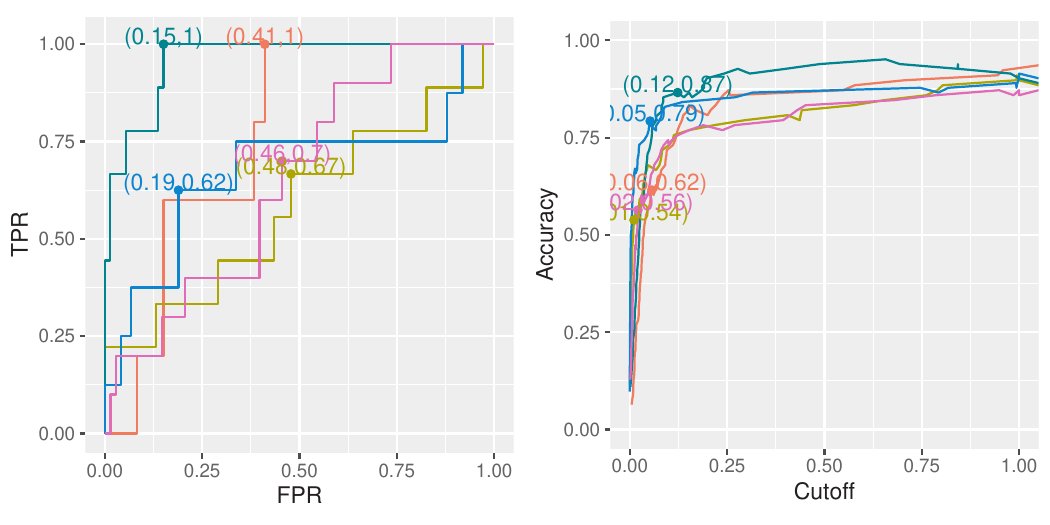}
    \caption{The maximum average AUC with classifier GLM~$\xrightarrow{}$~GLM, $w_\text{max} = 23$, $\mathcal{L}_1$ }
    \label{fig:fpr_tpr_glm_glm_te}
\end{figure}
\paragraph{GLM~$\xrightarrow{}$~GLM} The average value over the five iterations are FPR = 0.337, TPR = 0.798, cut-off = 0.053, Acc = 0.675. Here the cut-off value means that when the response of the second GLM model is greater than 0.053 then it is classified as dysgraphia, non-diagnosed otherwise.

\begin{figure}
    \centering
    \includegraphics[width = \textwidth]{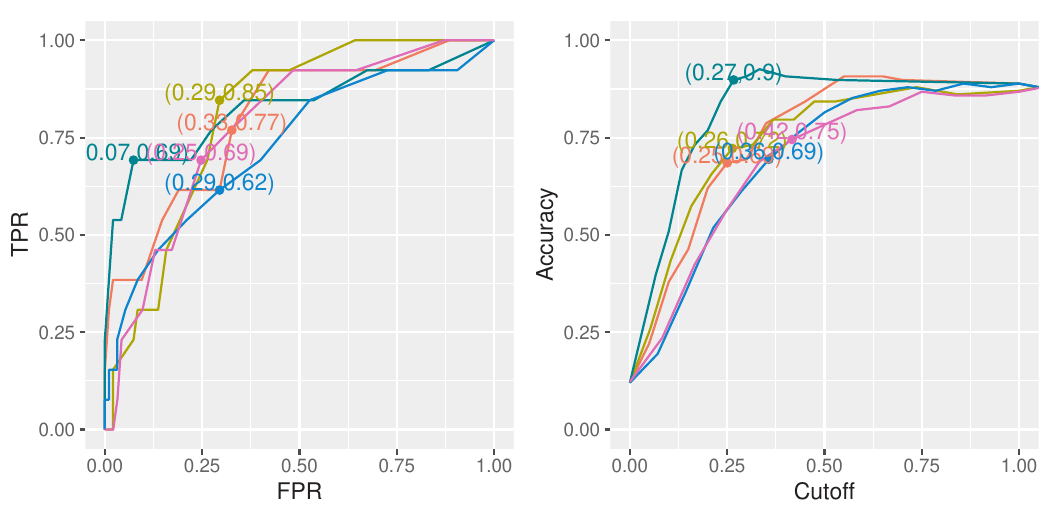}
    \caption{The maximum average AUC with combination RF~$\xrightarrow{}$~Counting, $w_\text{max}=35, \mathcal{L}_1, \alpha = 0.89$}
    \label{fig:fpr_tpr_rf_pi_te}
\end{figure}
\paragraph{RF~$\xrightarrow{}$~Counting} The average value over the five iterations are FPR = 0.247, TPR = 0.723, cut-off = 0.311, Acc = 0.749. Here the cut-off value means that when the number of symbols which are identified as written by someone with dygraphia by the first layer RF is greater than 11, then the individual is classified as dysgraphia.

\begin{figure}
\centering
    \includegraphics[width = \textwidth]{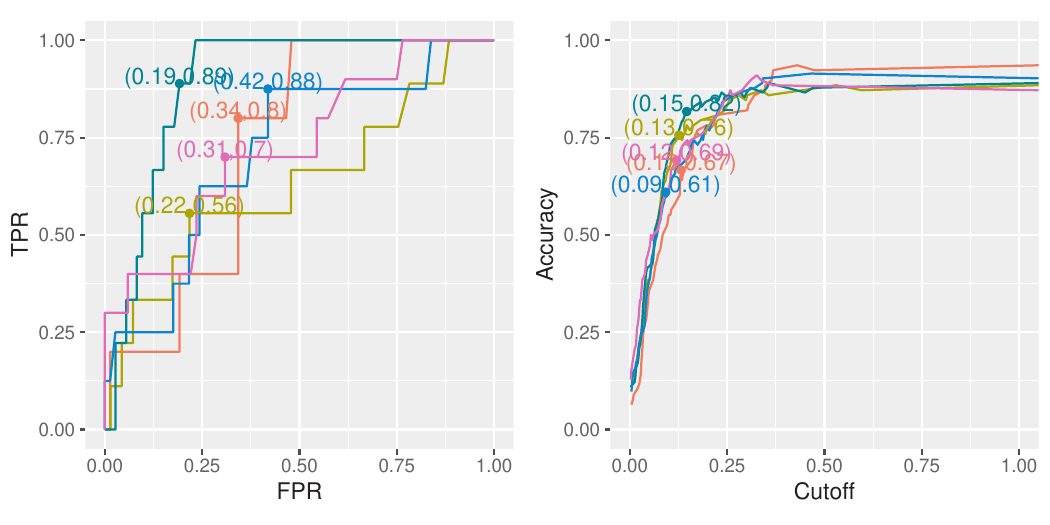}
    \caption{The maximum average AUC with combination RF~$\xrightarrow{}$~RF, $w_\text{max}=39, \mathcal{L}_{\infty}$}
\label{fig:fpr_tpr_rf_rf_te}
\end{figure}

\paragraph{RF~$\xrightarrow{}$~RF} The average value over the five iterations are FPR = 0.296, TPR = 0.764, cut-off = 0.123, Acc = 0.708. Here the cut-off value means that when 0.123 decision trees in the second layer RF vote for dysgraphia, then the individual is classified as dysgraphia.

\begin{figure}
\centering
    \includegraphics[width = \textwidth]{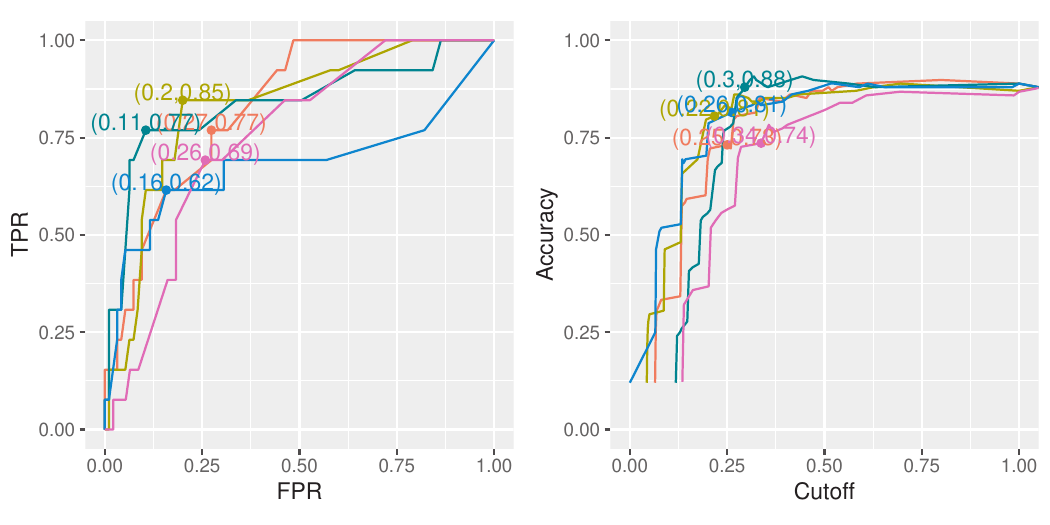}
    \caption{The maximum average AUC with classifier SVM~$\xrightarrow{}$~Counting, $w_\text{max}=27, \mathcal{L}_{\infty}$}
\label{fig:fpr_tpr_svm_count_te}
\end{figure}

\paragraph{SVM~$\xrightarrow{}$~Counting} The average value over the five iterations are FPR = 0.199, TPR = 0.738, cut-off = 0.272, Acc = 0.793. Here the cut-off value means that when the number of symbols which are identified as written by someone with dygraphia by the first layer RF is greater than 9 ($36\times0.272$), then the individual is classified as dysgraphia.

\subsection{SVM + SVM}\label{app:svm2}
\begin{figure}
    \centering
    \includegraphics[width = \textwidth]{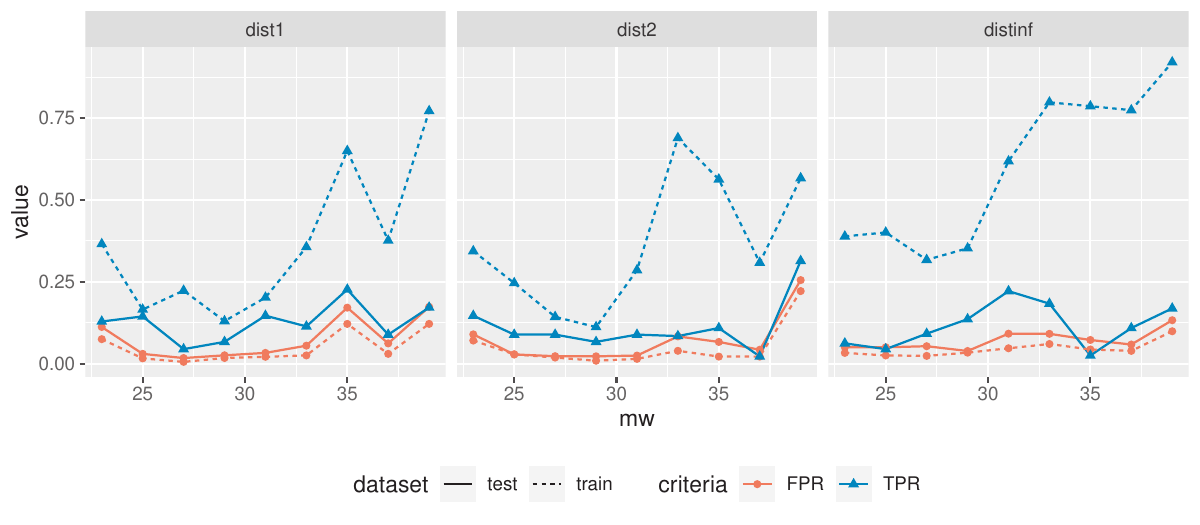}
    \caption{When probability is not used, the algorithm gives prediction 0 or 1. Consequently, the FPR and FPR of the results are obtained.}
    \label{fig:svm2}
\end{figure}
\begin{figure}
    \centering
    \includegraphics[width = \textwidth]{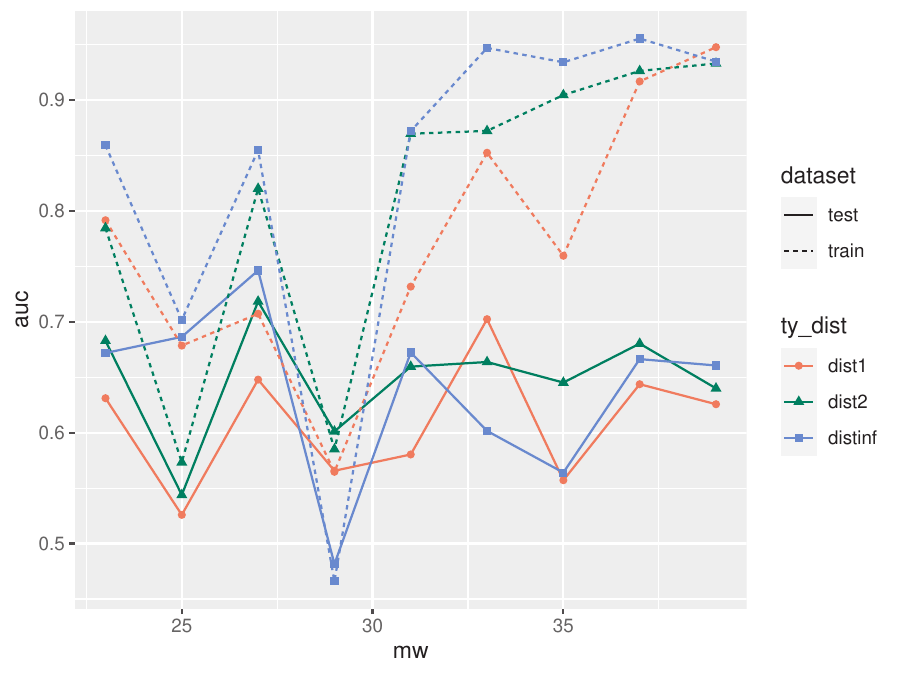}
    \caption{When probability is calculated}
    \label{fig:auc_svm_svm}
\end{figure}
\end{appendix} 

\bibliographystyle{ieeetr}
\bibliography{ref.bib}
\nocite{charles2004bhk}

\end{document}